\documentclass[trackchanges]{aastex701}
\usepackage{rotating}
\usepackage{adjustbox}
\usepackage{graphicx}
\usepackage{capt-of}  
\usepackage{multirow}

\newcommand{\Nv}{\ion{N}{5}}
\newcommand{\Niv}{\ion{N}{4}]}
\newcommand{\Niii}{\ion{N}{3}]}
\newcommand{\Civ}{\ion{C}{4}}
\newcommand{\Ciii}{\ion{C}{3}]}
\newcommand{\Mgii}{\ion{Mg}{2}}
\newcommand{\Feii}{\ion{Fe}{2}}
\newcommand{\Heii}{\ion{He}{2}}
\newcommand{\Aliii}{\ion{Al}{3}}
\newcommand{\Siiii}{\ion{Si}{3}]}
\newcommand{\Oiii}{\ion{O}{3}}

\def\IhepCAS{Key Laboratory for Particle Astrophysics, Institute of High Energy Physics, Chinese Academy of Sciences, 19B Yuquan Road, Beijing 100049, P. R. China}

\def\UCASast{School of Astronomy and Space Science, University of Chinese Academy of Sciences, 19A Yuquan Road, Beijing 100049, P. R. China}

\def\naocOptical{National Astronomical Observatories, Chinese Academy of Sciences (CAS), Beijing 100101, P. R. China}
\def\SpaceTechnology{Technology and Engineering Center for Space Utilization, Chinese Academy of Sciences, Beijing 100094, P. R. China}
\def\nanjing{School of Physics and Technology, Nanjing Normal University, No. 1, Wenyuan Road, Nanjing 210023, People’s Republic of China}

\begin{document}

\title{Nitrogen-Loud Quasars from the Dark Energy Spectroscopic Instrument. I. Sample Selection and Basic Properties}

\author[0009-0005-4152-2088]{Shuo Zhai}
\email{}
\affiliation{\naocOptical}

\author[0000-0001-9457-0589]{Wei-Jian Guo}
\email{}
\affiliation{\naocOptical}

\author[0000-0003-4280-7673]{Yong-Jie Chen}
\email{}
\affiliation{\SpaceTechnology}

\author[]{Zhi-Qiang Chen}
\email{}
\affiliation{\nanjing}

\author[0000-0002-0389-9264]{Haining Li}
\email{}
\affiliation{\naocOptical}
\affiliation{\UCASast}––

\author[0000-0001-9449-9268]{Jian-Min Wang}
\email{}
\affiliation{\IhepCAS}
\affiliation{\naocOptical}
\affiliation{\UCASast}


\author[0000-0002-8980-945X]{Gang Zhao}
\email{}
\affiliation{\naocOptical}
\affiliation{\UCASast}

\begin{abstract}

We present the largest sample to date of nitrogen-loud (N-loud) quasars
with strong broad \Niv~$\lambda1486$ and/or
\Niii~$\lambda1750$ emission lines over the redshift range
$1.6<z<4.3$, selected from the Dark Energy Spectroscopic Instrument
(DESI) Data Release 1. The final sample contains {\color{black}2,198} N-loud
quasars, corresponding to about {\color{black}$1.3\%$} of the parent quasar sample. The
$L_{1450}$ distribution of the N-loud quasars is broadly similar to that
of the DESI parent sample, but their redshift distribution is distinct,
with a stronger concentration around $z\sim2.5$--3. Their composite
spectrum displays a broadly similar UV continuum shape to that
of the parent quasars, while showing significantly enhanced broad
nitrogen emission features, including \Nv, \Niv,
and \Niii. Other metal emission features also show a moderate
enhancement. Relative to a control sample matched in redshift and UV
continuum luminosity, the N-loud quasars show systematically narrower
broad \Civ\ and \Mgii\ emission lines, lower
single-epoch virial black hole masses, and higher Eddington ratios,
suggesting that N-loud quasars may preferentially appear during a
relatively rapid black hole accretion phase. The radio-loud fraction is
{\color{black}$9.8\%$}, with the highest fraction among objects exhibiting both
\Niii\ and \Niv\ emission. The catalog provides a
statistical baseline for future studies of nitrogen enhancement and its
physical origin.

\end{abstract}


\section{Introduction}
\label{sec:intro}

Nitrogen-loud (N-loud) quasars,\footnote{Quasars with strong broad
\Niii\ and/or \Niv\ emission lines have historically
been referred to as both N-rich and N-loud quasars. In this paper, we
adopt the term N-loud quasars consistently, while retaining the original
terminology when referring to previous studies.} characterized by
unusually strong broad nitrogen emission lines, form a rare but
recognizable subclass of quasars. The best-known example is Q0353$-$383,
whose spectrum exhibits prominent \Niv~$\lambda1486$,
\Niii~$\lambda1750$, and \Nv~$\lambda1240$ emission lines
\citep{Osmer1980}. Such objects have long been used to investigate
chemical enrichment in the broad line region (BLR), since strong
nitrogen emission is generally thought to be associated with anomalous
chemical abundances, and may in particular reflect enhanced nitrogen
abundance or high metallicity
\citep{Baldwin2003,Batra2014,Matsuoka2017}. In recent years, with the
rapid progress of James Webb Space Telescope observations of chemically peculiar sources at
high redshift, strong nitrogen emission and its physical origin have
again attracted broad attention \citep{Bunker2023,Ji2024,Ji2026,Isobe2025}.
Although these high-redshift sources are not necessarily identical to the
classical N-loud quasars, a well-defined low- to intermediate-redshift
N-loud quasar sample can provide an important reference for understanding
the strong-nitrogen phenomenon and a statistical basis for future
abundance analyses based on broad emission lines.

Systematic searches based on the Sloan Digital Sky Survey (SDSS)
gradually established N-loud quasars as a rare spectroscopic subclass.
\citet{Bentz2004b} first identified 16 candidate nitrogen-enriched
quasars in the SDSS Early Data Release, initiating a systematic search
for such objects. \citet{Bentz2004a} subsequently reported 20 N-rich
quasars in the SDSS First Data Release, corresponding to about 0.4\% of
the sample. \citet{Jiang2008} further identified 293 quasars with strong
\Niv\ or \Niii\ emission over the redshift range
\(1.7<z<4.0\), providing the first statistically meaningful sample of
N-loud quasars. \citet{Batra2014} then carried out BLR metallicity
analyses for 43 objects with the strongest \Niv\ and
\Niii\ emission, finding that these sources tend to have high BLR
metallicities and relatively low black hole masses. In addition to the
SDSS-based optical studies, \citet{Matsuoka2017} obtained near-infrared
spectroscopy for 12 N-loud quasars and used rest-frame optical emission
lines to further investigate their chemical properties and accretion
states. These studies have revealed several important features of
N-loud quasars, but the existing samples remain limited in size, data
homogeneity, and selection strategy. In particular, the incidence of
N-loud quasars relative to the parent quasar population, their redshift
and luminosity distributions, their broad line and black hole properties,
and their radio properties still require systematic characterization with
a larger and more uniformly selected sample. Such a statistical baseline
is also needed for further assessing the physical origin of the strong
nitrogen emission.

In addition to these statistical limitations, the abundance pattern
underlying the strong nitrogen emission remains ambiguous. A common
interpretation links strong \Nv, \Niv, and \Niii\ emission to high
metallicity, because nitrogen is expected to behave mainly as a secondary
element at high metallicity, with \({\rm N/O}\) and \({\rm N/C}\)
increasing with the overall metal abundance. In this framework,
\citet{Baldwin2003} argued that the very high N/O inferred for the
extreme N-rich quasar Q0353$-$383 implies very high overall
metallicity, and \citet{Batra2014} reached a similar conclusion from
broad line analyses. However, strong nitrogen lines do not uniquely
require an overall metallicity increase. Using near-infrared rest-frame
optical spectra, \citet{Matsuoka2017} found that some N-loud quasars have
[\Oiii]~\(\lambda5007\) strengths comparable to those of normal
quasars, rather than the strongly suppressed [\Oiii] emission expected
for extremely metal-rich narrow line region gas. This suggests that strong broad
nitrogen lines do not necessarily indicate extremely high overall
metallicity, but may instead reflect nitrogen enhancement relative to
oxygen and other elements. Therefore, strong \Niii\ or \Niv\ emission
alone cannot distinguish between high overall metallicity and selective
nitrogen enhancement; this requires multi-line abundance diagnostics and
dedicated photoionization modeling based on a uniformly selected sample.

The large, homogeneous spectroscopic database and broad wavelength
coverage of Dark Energy Spectroscopic Instrument (DESI) allow us to search for broad \Niv~\(\lambda1486\) and
\Niii~\(\lambda1750\) emission simultaneously over a wide redshift range,
thereby enabling a new statistical study of N-loud quasars \citep{DESI2016a}. This paper is
the first in a series on DESI N-loud quasars, and focuses on constructing
the sample and characterizing its basic statistical properties. 
We construct an N-loud quasar catalog based on DESI Data Release 1 (DR1) and
measure their redshifts, UV continuum luminosities, equivalent width (EW) and
FWHM of the major broad emission lines, single-epoch virial black hole
masses, and Eddington ratios. We further
characterize the radio properties of the sample through cross-matching
with radio surveys. In the next paper of this series, we will use
multiple UV emission line ratios and photoionization models to
examine the abundance pattern of the sample, testing whether the strong
nitrogen emission reflects an overall increase in metallicity or
selective nitrogen enhancement relative to other elements.

The paper is organized as follows. In Section~\ref{sec:selection}, we
describe the DESI DR1 data, the construction of the parent sample, the
N-loud quasar selection procedure, and the spectral fitting used to
measure the relevant emission line properties. In Section~\ref{sec:result},
we present the basic properties of the selected N-loud quasars and
compare their spectral, black hole, and radio properties with control
samples. Our main conclusions are summarized in Section~\ref{sec:sum}.
Throughout this paper, we adopt a flat \(\Lambda\)CDM cosmology with
\(H_0 = 67.4~\mathrm{km~s^{-1}~Mpc^{-1}}\),
\(\Omega_{\rm m} = 0.315\), and \(\Omega_\Lambda = 0.685\)
\citep{Planck2020}.

\section{Data and Sample Selection}\label{sec:selection}

\subsection{DESI}\label{sec:desi}

DESI is a Stage-IV wide-area
ground-based multi-object spectroscopic survey mounted on the
NOIRLab 4 m Mayall Telescope at Kitt Peak National Observatory
\citep{Levi2013,DESI2016a,DESI2016b,DESI2022,Schlafly2023}.
DESI is equipped with 5000 fibers and provides continuous wavelength
coverage over 3600--9800\,\AA\ through three-arm spectrographs, with
typical spectral resolutions of
\(\lambda/\Delta\lambda \sim 2100\), 3200, and 4100 for the three
channels, respectively \citep{Miller2024,Silber2023,DESI2016b}.
We use spectra from DESI DR1, which includes observations obtained
between 2021 May and 2022 June. The catalog spectra are coadded from
multiple tiles or exposures, and the object classifications and redshift
measurements are provided by the Redrock spectral template-fitting
pipeline \citep{Brodzeller2023}. 

We adopt the Redrock-based quasar classifications in DR1 as the starting
point for constructing the parent sample, from which we further identify
quasar candidates with unusually strong nitrogen emission lines. 
All spectra are corrected for Galactic extinction using the
extinction curve of \citet{Fitzpatrick1999} with \(R_V=3.1\), and are
then shifted to the rest frame for the subsequent sample selection and
emission line measurements.

\subsection{Sample Selection}\label{sec:criteria}

Our parent sample was constructed from the DESI DR1 catalog by selecting
sources classified as quasars (\texttt{SPECTYPE=QSO}) and requiring
\texttt{ZWARN=0}, thereby retaining only objects with reliable redshift
measurements \citep{DESI2025}. {\color{black}We further restricted the sample to
\(1.6 \leq z \leq 4.3\), over which the DESI spectra share a common
rest-frame wavelength coverage of approximately 1385--1850\,\AA,
encompassing both \Niv\; and \Niii.
To ensure sufficient continuum quality for the subsequent spectral fitting
and emission-line measurements, we required a continuum \(\mathrm{S/N}>5\) in the
rest-frame 1675--1725\,\AA\ window. This window is relatively free of
strong emission features and provides a consistent measure of the UV
continuum quality across the sample. The continuum S/N is defined as the
ratio of the mean flux density to the mean flux uncertainty within this
window.} These criteria define the parent sample used
for the subsequent selection, which contains 168,500 quasars.

\begin{figure*}[htbp]
    \centering
    \includegraphics[width=\textwidth]{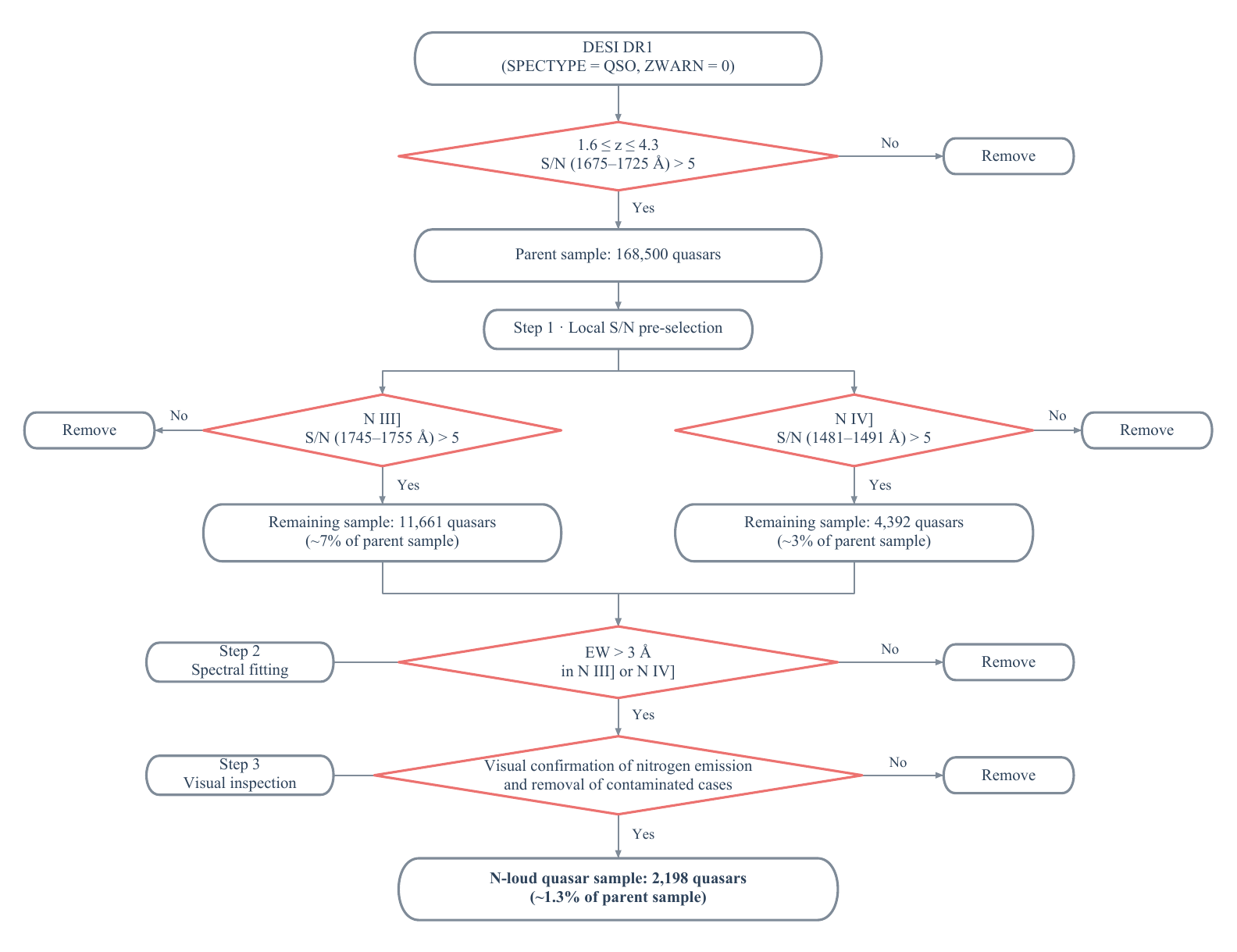}
    \caption{\footnotesize
    {\color{black}
    Selection flowchart for the DESI DR1 N-loud quasar sample. Starting from DESI DR1 quasars with \texttt{SPECTYPE=QSO} and \texttt{ZWARN=0}, we first require $1.6 \le z \le 4.3$ and $\mathrm{S/N}(1675\text{--}1725,\mathrm{\AA}) > 5$, yielding a parent sample of 168,500 quasars. We then apply a local S/N pre-selection around the two nitrogen lines, requiring $\mathrm{S/N}(1745\text{--}1755\,\mathrm{\AA}) > 5$ for \Niii\ or $\mathrm{S/N}(1481\text{--}1491\,\mathrm{\AA}) > 5$ for \Niv, leaving 11,661 ($\sim$7\% of the parent sample) and 4,392 ($\sim$3\%) objects, respectively. The pre-selected candidates are then subjected to spectral fitting, and objects with $\mathrm{EW} > 3,\mathrm{\AA}$ in either \Niii\ or \Niv\ are retained. Finally, visual inspection is performed to confirm genuine nitrogen emission and remove contaminated cases. The final N-loud quasar sample contains 2,198 quasars, corresponding to $\sim$1.3\% of the parent sample.}
    }
    \label{fig:selection}
\end{figure*}

We then identified N-loud quasar candidates from this parent sample in
three steps, following the procedure illustrated in
Figure~\ref{fig:selection}.
First, before performing the full spectral fitting, we used the local
line S/N around  {\color{black}
both \Niii\; and \Niv\; as pre-selection metrics. For \Niii,} the local continuum
was estimated from two sidebands on either side of \Niii,
1730--1740\,\AA\ and 1760--1770\,\AA, which are relatively free of
strong emission features. We then computed the continuum-subtracted
integrated flux within 1745--1755\,\AA\ and divided it by the propagated
flux uncertainty in the same window to define the local line S/N. We
required this local line \(\mathrm{S/N}>5\) to remove spectra with no
significant local excess around \Niii. After this pre-selection, the
sample was reduced to 11,661 quasars, corresponding to about 7\% of the
parent sample. 
{\color{black}
For \Niv, the local continuum was estimated from two sidebands at 1466--1476\,\AA\ and 1496--1506\,\AA. We then computed the continuum-subtracted integrated flux within 1481--1491\,\AA\ and required the corresponding local line \(\mathrm{S/N}>5\), yielding 4,392 quasars, or about 3\% of the parent sample. These local S/N criteria are used only as an efficient pre-selection step to remove spectra without significant local excess near either nitrogen line.}
Second, we performed spectral fitting for the {sources passing either of the above local S/N pre-selections}
and measured the EW of \Niii\; and \Niv; the fitting procedure
and line measurements are described in Section~\ref{sec:fitting}. We
retained objects with rest-frame \(\text{EW}>3\,\mathrm{\AA}\) in either \Niii\; or
\Niv. This threshold was adopted for two reasons: (1) for spectra passing
the above local line S/N pre-selection, a nitrogen line with
\(\text{EW}=3\,\mathrm{\AA}\) corresponds to a detection significance greater than
\(5\sigma\), thus removing most spurious features caused by noise
fluctuations; and (2) this threshold is consistent with the criteria used
in previous N-loud quasar searches, facilitating direct comparison
between the DESI sample and earlier SDSS-based samples
\citep{Bentz2004a,Jiang2008}.
Finally, all fitted spectra were visually inspected to confirm the
presence of nitrogen emission and to remove spurious cases caused by
absorption features, noise spikes, or poor local continuum subtraction.
The final sample contains {\color{black}2,198} N-loud quasars, corresponding to about
{\color{black}1.3\%} of the parent sample; the resulting catalog is presented in
Table~\ref{tab:Nloud}.

\subsection{Spectral Fitting and Line Measurements}\label{sec:fitting}

To obtain uniform line measurements for the N-loud quasar selection and
for the subsequent statistical characterization, we performed spectral
fitting for the {\color{black}16,053} sources that passed the local line S/N
pre-selection described in Section~\ref{sec:criteria}. We fitted the spectra with DASpec \citep{Du2018}, a spectral decomposition code that employs
the Levenberg--Marquardt algorithm to minimize $\chi^2$ and fits all spectral components simultaneously, over the rest-frame wavelength range from 1450\,\AA\ to the maximum rest-frame wavelength allowed by the DESI
observed-frame coverage. Based on whether the \Mgii\; and \Feii\;
regions remain within the DESI observed-frame wavelength coverage, we
divided the fitted sources into three redshift intervals: for sources at
\(z<2.37\), the fitting window includes both the \Mgii\; and \Feii\;
emission regions; for sources with \(2.37<z<2.62\), the fitting includes
the \Feii\; complex but excludes \Mgii; and for sources at
\(2.62<z<4.3\), neither \Mgii\; nor \Feii\; is included.

The pseudo-continuum and emission lines were fitted simultaneously. The
continuum was modeled as a power law. Within the 1450--2000\,\AA\ region,
the major emission lines \Niv, \Civ, \Niii, \Heii, and \Ciii\; were
fitted with Gaussian profiles. Specifically, \Niii, \Niv, \Heii, and
\Ciii\; were fitted with single Gaussian profiles; for \Civ, both
single- and double-Gaussian models were considered, and the model with
the lower reduced \(\chi^2\) was adopted. The \Ciii\; region may be
affected by nearby weak emission features, such as
\Aliii\(\lambda1857\) and \Siiii\(\lambda1892\), but we did not attempt
a detailed decomposition of these weak components. Because the N-loud
selection is based only on \Niii\; and \Niv, this simplified treatment of
the \Ciii\; complex does not affect the sample definition. For sources
whose fitting window covers the \Feii\; region, the \Feii\; emission was
modeled using a broadened and scaled version of the
\citet{Vestergaard2001} template; for sources whose fitting window covers
\Mgii, the \Mgii\; line was fitted with a single Gaussian profile.

\begin{figure*}[htbp]
    \centering
    \includegraphics[width=0.5\textwidth]{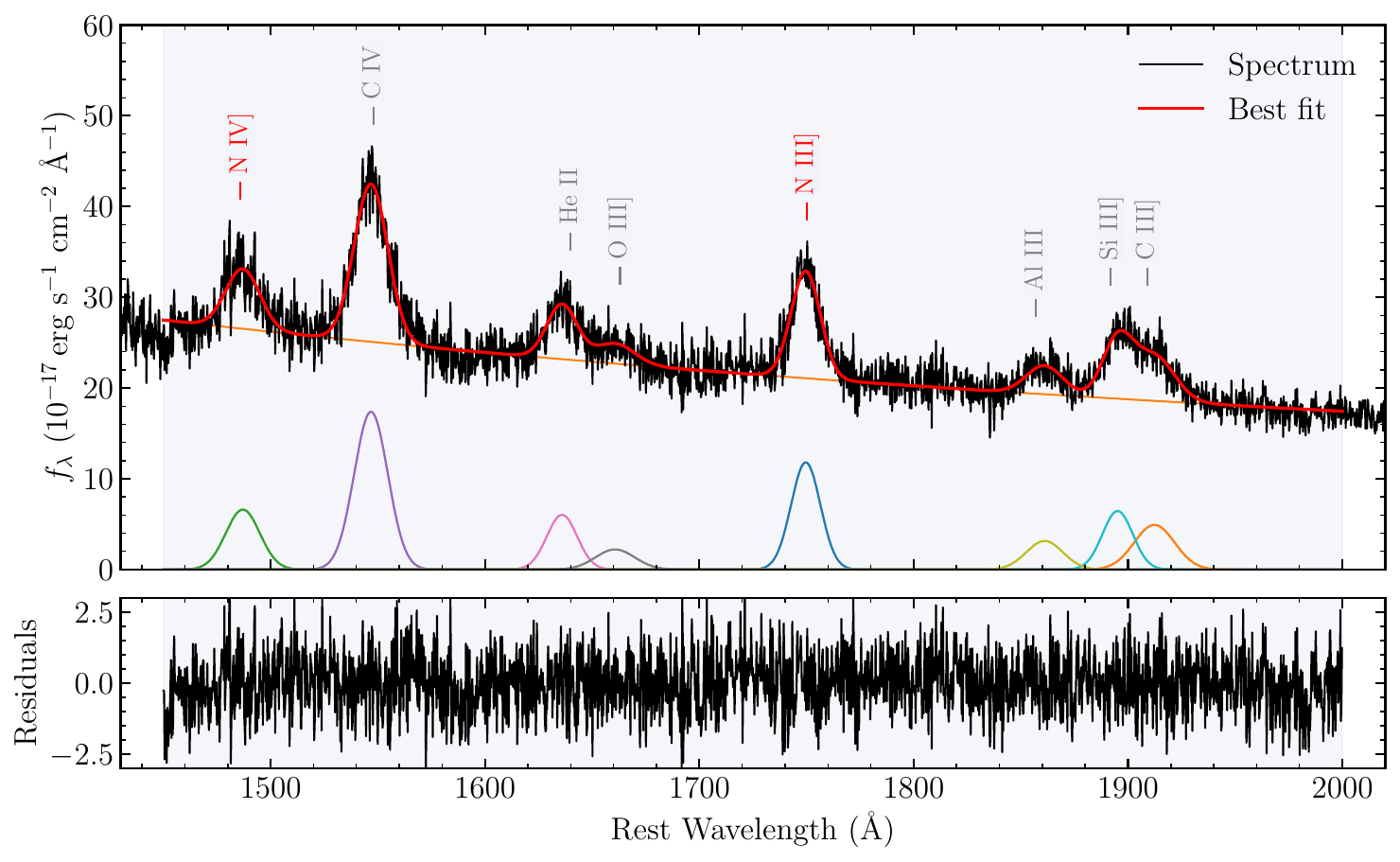}
    \caption{\footnotesize
    Spectral decomposition of an example N-loud quasar. The top panel shows the fitting result, and the bottom panel shows the residuals. The fitting window is indicated by the shaded region. The observed rest-frame spectrum is shown in black, and the total best-fit model is shown in red. The orange line shows the fitted continuum, and the individual emission line components are marked with different colored curves. The broad nitrogen features \Niv\; and \Niii, which define the N-loud selection, are highlighted in red.}
    \label{fig:fitting}
\end{figure*}

The FWHM and EW are measured from the best fits. The EW of \Niii\; and \Niv\; were used
for the EW-based sample selection described in
Section~\ref{sec:criteria}. We then visually inspected all fitted spectra
that passed the EW threshold to confirm the presence of nitrogen emission
and to remove cases in which the relevant nitrogen features were
severely affected by absorption features, noise spikes, or poor local
continuum subtraction. An example of the spectral decomposition is shown
in Figure~\ref{fig:fitting}.

\section{results}\label{sec:result}

Applying the selection procedure described in Section~\ref{sec:selection},
we identify {\color{black}2,198} N-loud quasars from the DESI DR1 quasar catalog over
the redshift range \(1.6<z<4.3\), corresponding to about {\color{black}1.3\%} of the
parent sample. The resulting catalog is presented in
Table~\ref{tab:Nloud}, including the source identifiers, coordinates,
redshifts, continuum properties, major broad line measurements, and
additional quantities derived from the spectral fitting.

In this section, we first summarize the sample composition and its
distribution in redshift and luminosity
(Section~\ref{sec:sample}). We then examine the spectral
properties of the N-loud quasars, including their nitrogen line
properties, composite spectrum, and broad line properties relative to
normal quasars (Section~\ref{sec:lines}). Finally, we
compare their single-epoch virial black hole masses and Eddington ratios
with those of normal quasars (Section~\ref{sec:BHmass}) and summarize
their radio properties (Section~\ref{sec:radio}).

\subsection{Basic Statistics}\label{sec:sample}

\begin{figure*}[htbp]
    \centering
    \includegraphics[width=\textwidth]{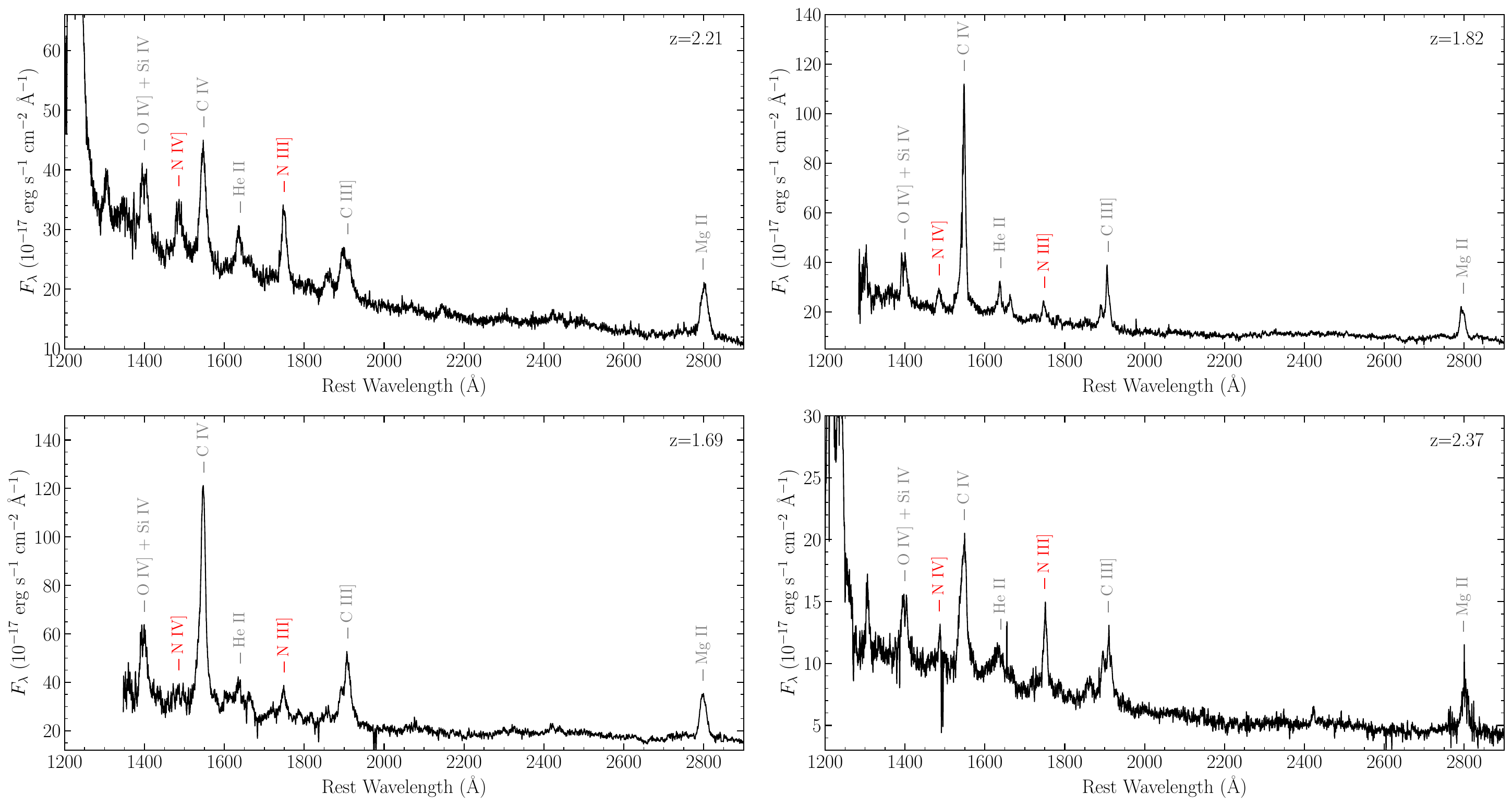}
    \caption{\footnotesize 
    Example spectra of four N-loud quasars. {\color{black}The black lines show the spectra smoothed with a 5-pixel boxcar.} The black dotted lines mark the positions of the \Niv\;and \Niii\;lines. The two upper panels show quasars with strong \Niv\;and \Niii\;lines, and the two lower panels show quasars with strong \Niii\;lines.}
    \label{fig:spec}
\end{figure*}

{\color{black}
The final sample consists of 2,198 N-loud quasars with different combinations of prominent nitrogen emission. Among them, 303 satisfy both the \Niii\; and \Niv\; selection criteria, 1,669 are selected only through strong \Niii, and 226 only through strong \Niv.}
Figure~\ref{fig:spec}
shows four examples of the selected quasars. The first two spectra
exhibit strong \Niii\; and \Niv\; emission simultaneously, whereas the
latter two are dominated mainly by strong \Niii\; emission. Although the
continuum shapes and the strengths of other emission features vary from
object to object, prominent nitrogen emission is the defining feature of
all selected sources.

\begin{figure*}[htbp]
    \centering
    \includegraphics[width=0.5\textwidth]{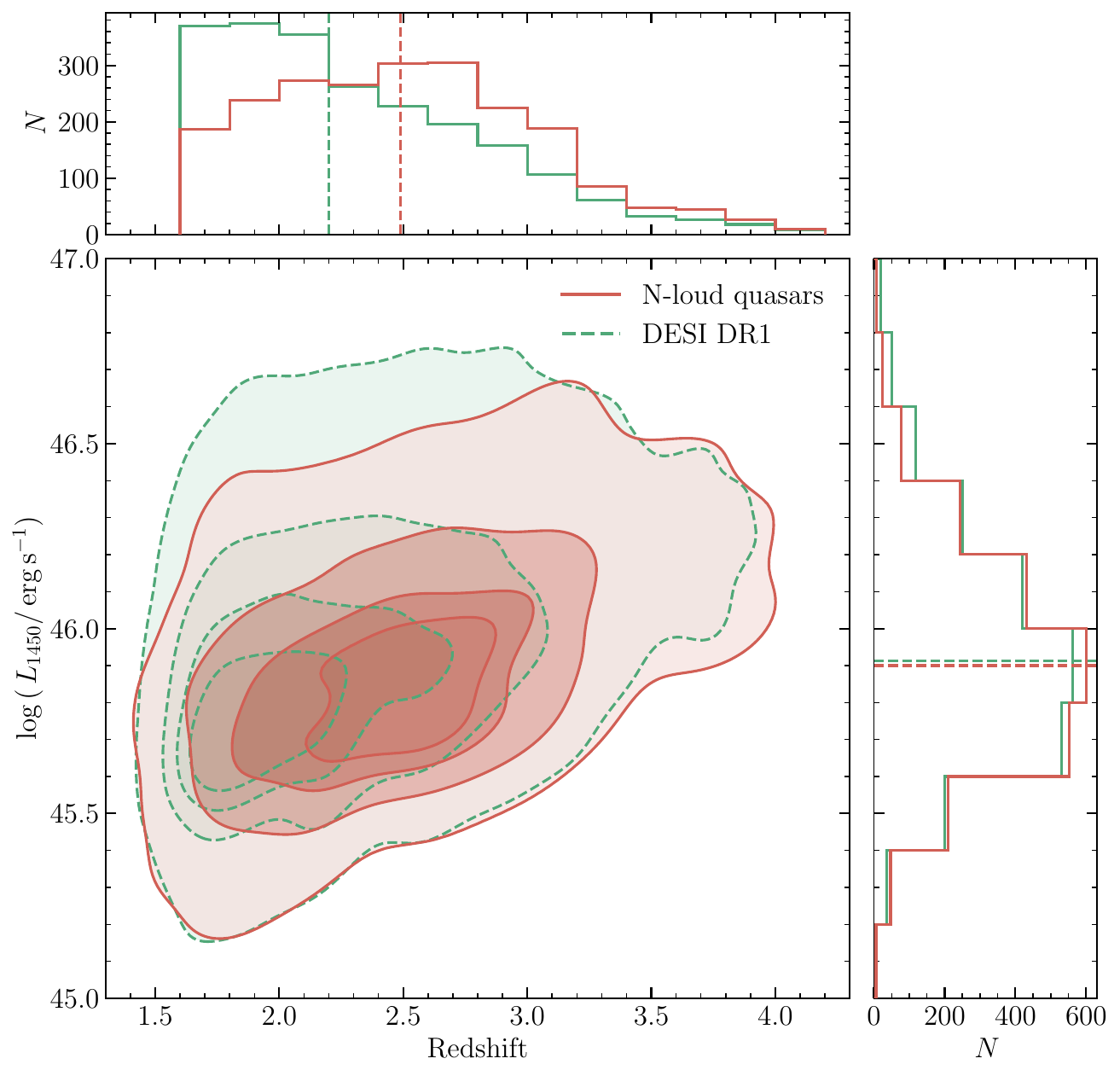}
    \caption{\footnotesize Redshift versus rest-frame 1450~\AA{} luminosity distribution for the N-loud quasars and the DESI DR1 parent quasar sample. The 20\%, 40\%, 68\%, and 95\% percentile density contours are shown, with shaded regions indicating the corresponding density levels. Red solid contours represent the N-loud quasars, while green dashed contours represent the DESI DR1 parent quasar sample. The top and right panels show the corresponding one-dimensional distributions of redshift and $\log L_{1450}$, respectively, with dashed lines indicating the median values. {\color{black}For comparison, the DESI DR1 quasar histograms are scaled to match the number of N-loud quasars.}}
    \label{fig:contour_z_L1450}
\end{figure*}

Figure~\ref{fig:contour_z_L1450} first compares the location of the N-loud
quasars and the DESI DR1 parent quasar sample in the
\(z\)--\(L_{1450}\) plane\footnote{
We estimated the monochromatic luminosity $\lambda L_\lambda(1450\,\text{\AA})$ from the median rest-frame continuum flux over 1440--1460\,\text{\AA}, using $\lambda L_\lambda(1450\,\text{\AA}) = 4\pi D_L^2 \lambda f_\lambda(1450\,\text{\AA})$. For simplicity, we denote $\lambda L_\lambda(1450\,\text{\AA})$ as $L_{1450}$ throughout this paper.
}. The two samples occupy broadly similar regions
in this parameter space, and their one-dimensional \(L_{1450}\)
distributions are nearly identical, indicating that the selected
N-loud quasars are not confined to a particular luminosity range.
However, their one-dimensional redshift distributions differ, with the
N-loud quasars more concentrated around \(z\sim2.5\)--3.

\begin{figure*}[htbp]
    \centering
    \includegraphics[width=0.5\textwidth]{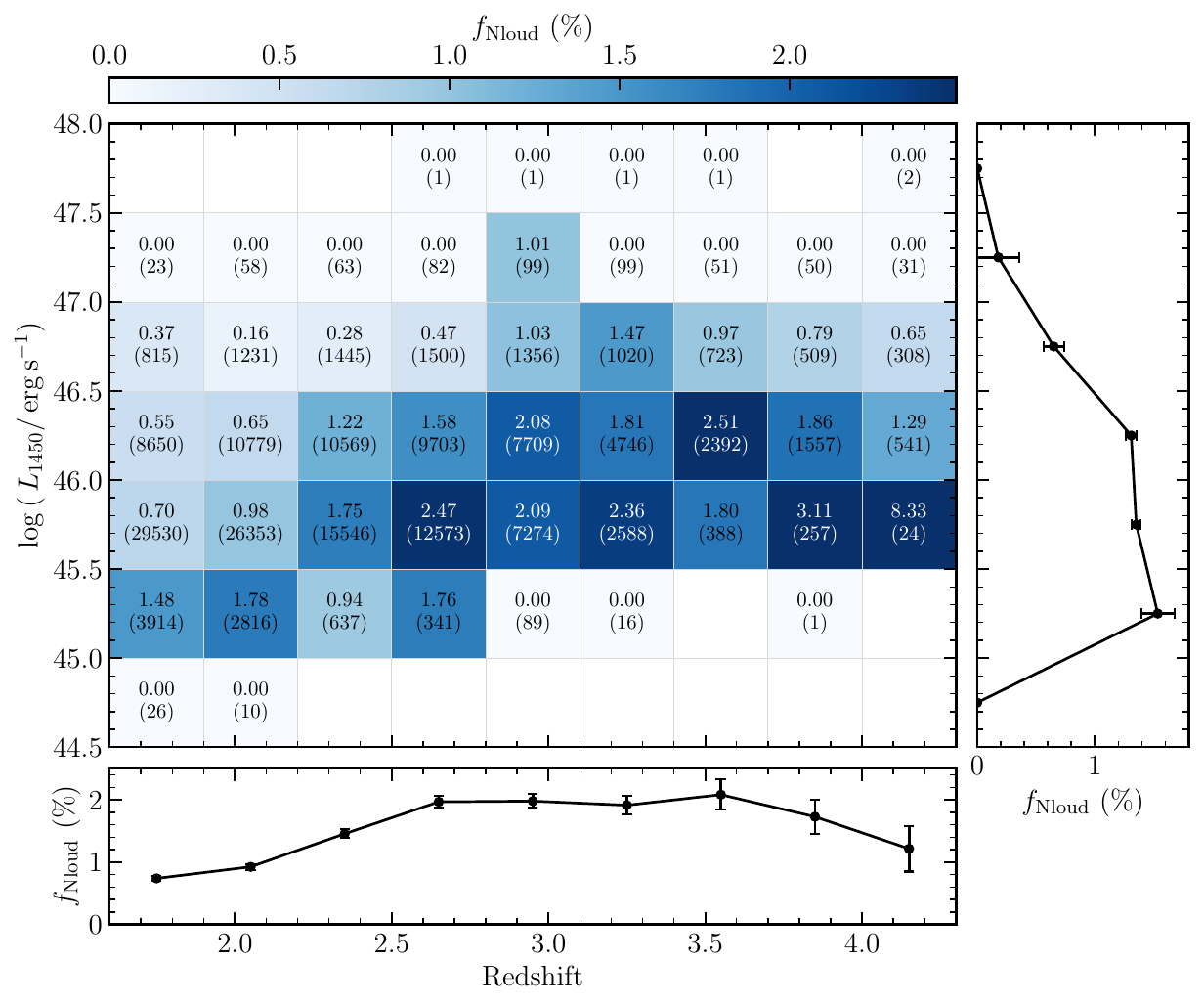}
    \caption{\footnotesize
    Observed fraction of N-loud quasars as functions of redshift and rest-frame 1450~\AA{} luminosity. The numbers in parentheses are the total numbers of quasars in each bin, and only bins with N-loud quasar detections are shown. The bottom and right panels show the corresponding one-dimensional fractions as functions of redshift and $\log L_{1450}$, respectively, with error bars indicating the statistical uncertainties.}
    \label{fig:z_L1450_grid}
\end{figure*}

Figure~\ref{fig:z_L1450_grid} then quantifies the observed fraction of
N-loud quasars across the same \(z\)--\(L_{1450}\) parameter space.
Consistent with the redshift-distribution difference shown in
Figure~\ref{fig:contour_z_L1450}, the observed N-loud fraction is not uniform
with redshift: it increases from \(z\sim1.6\) to \(z\sim3\), reaches a
peak of about 2\%, and then declines toward higher redshift. By contrast,
the dependence of the N-loud fraction on \(L_{1450}\) is weaker and
should be interpreted with caution, especially in bins with small
parent-sample numbers. 
{\color{black}
Because the detectability of weak nitrogen emission may depend on spectral quality, Appendix~\ref{sec:SN_redshift} examines the effect of the continuum S/N selection on the observed redshift dependence. The enhancement around $z\sim2.5$--3 remains present under stricter S/N thresholds (Figure~\ref{fig:SNR_robustness}), indicating that this feature is not primarily driven by the adopted continuum S/N cut.
}

%

The redshift-distribution difference between the DESI N-loud quasars and
the DESI parent sample contrasts with the result of \citet{Jiang2008},
whose N-rich quasar sample more closely follows the redshift distribution
of the SDSS parent sample over a similar redshift range. This difference
may partly reflect the different parent-sample definitions.
\citet{Jiang2008} searched SDSS DR5 quasars with \(1.7<z<4.0\) and
\(i<20.1\), so their sample is closer to a magnitude-limited sample. In
contrast, our DESI parent sample is defined by \(1.6<z<4.3\) and further
requires continuum \(\mathrm{S/N}>5\) in the rest-frame
1675--1725\,\AA\ window, and is more directly affected by
spectral-quality requirements. 
{\color{black}
These two selections therefore probe somewhat different parent quasar populations, even over a similar redshift range. Although the redshift enhancement within our DESI sample is robust against reasonable changes in the continuum S/N threshold, the difference between the DESI and SDSS N-loud redshift distributions should not by itself be interpreted as evidence for intrinsic cosmic evolution without accounting for the different parent-sample selection functions.
}

\subsection{Spectral Properties}\label{sec:lines}

\begin{figure*}[htbp]
    \centering
    \includegraphics[width=0.5\textwidth]{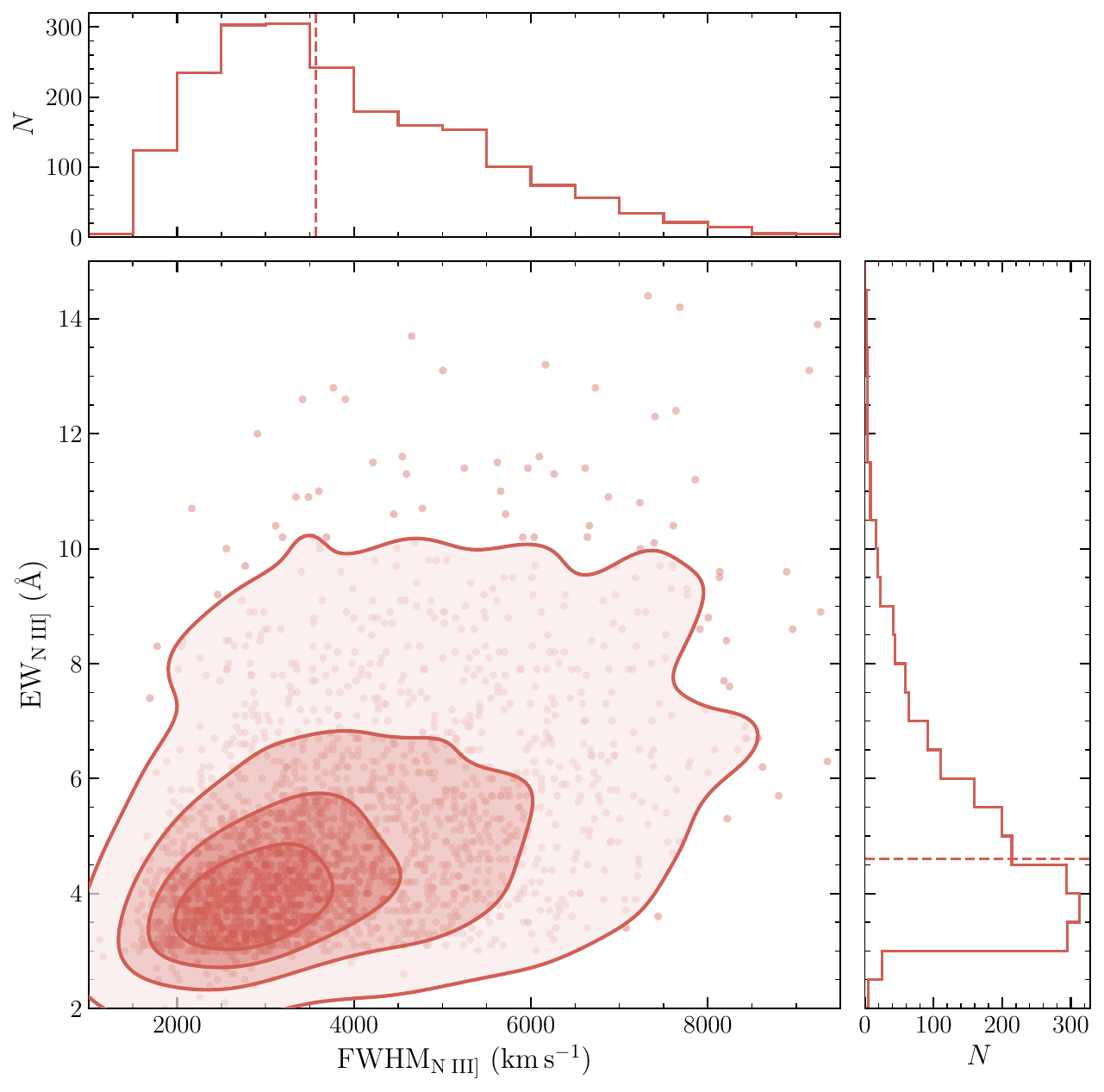}
    \caption{\footnotesize
    Rest-frame EW versus FWHM distribution of \Niii\;for the N-loud quasars. Red points show individual objects. The 20\%, 40\%, 68\%, and 95\% contours and the corresponding density maps are shown. The top and right panels show the corresponding one-dimensional distributions of FWHM and EW, respectively, with dashed lines indicating the median values.}
    \label{fig:contour_Niii}
\end{figure*}

{\color{black}We first examine the nitrogen-emission properties of individual N-loud quasars.}
Figure~\ref{fig:contour_Niii} shows the distribution of the \Niii\; EW
and FWHM in the N-loud sample, providing a statistical characterization
of the nitrogen line strength and width. We use \Niii\; rather than
\Niv\; as the main statistical line because \Niii\; is detected in
{\color{black}90\%} of the final sample and lies in a relatively clean spectral region
with less contamination from other strong emission lines. In contrast,
\Niv\; lies close to \Civ\; and is more easily affected by the broad
\Civ\; profile and local fitting uncertainties. Therefore, \Niii\;
serves as the primary statistical tracer of nitrogen emission in this
sample. The \Niii\; EW distribution peaks at about 4.0~\AA, with a
median of about 4.6~\AA. The FWHM distribution peaks at about
\(3000~\mathrm{km~s^{-1}}\), with a median of about
\(3600~\mathrm{km~s^{-1}}\). Most objects are concentrated near these
peak values, with the distributions extending toward larger EW and FWHM.
Compared with \citet{Jiang2008}, the EW distribution is broadly similar,
while in their sample both the peak and median of the \Niii\; FWHM
distribution are near \(4000~\mathrm{km~s^{-1}}\), somewhat higher than
in our sample.
\begin{figure*}[htbp]
    \centering
    \includegraphics[width=0.49\textwidth]{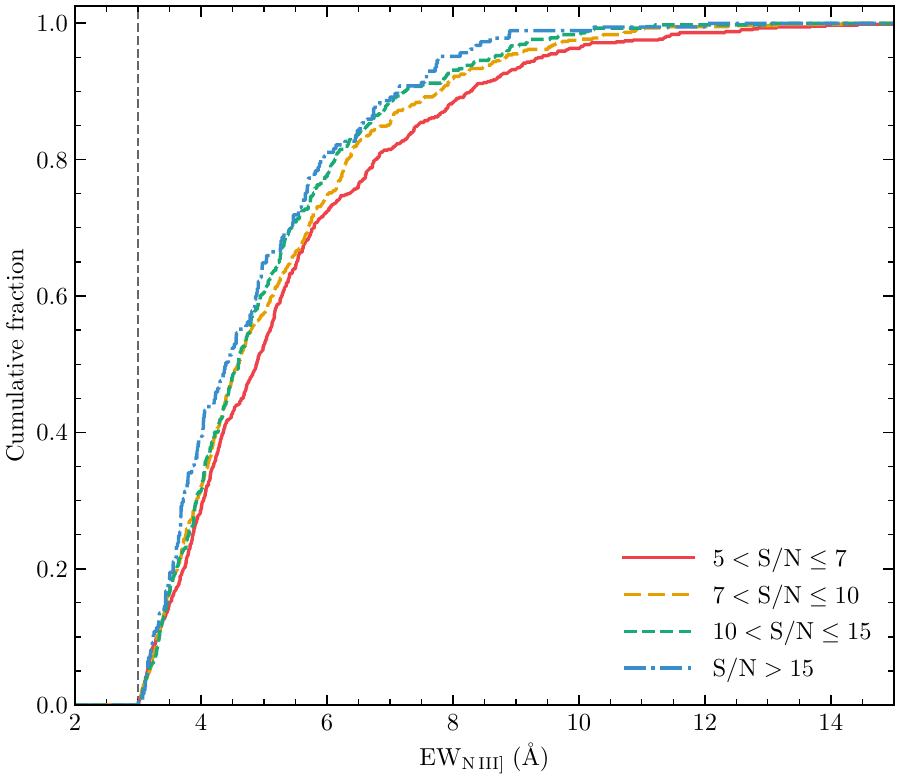}
    \caption{\footnotesize
    {\color{black}Cumulative distributions of rest-frame \Niii\; EW for the N-loud quasars in four continuum S/N bins. Red solid, orange dashed, green dashed, and blue dash-dotted curves show the distributions for $5<\mathrm{S/N}\leq7$, $7<\mathrm{S/N}\leq10$, $10<\mathrm{S/N}\leq15$, and $\mathrm{S/N}>15$, respectively. The continuum S/N is measured in the rest-frame 1675--1725~\AA{} window. The vertical dashed line marks the EW selection threshold of 3~\AA.}}
    \label{fig:niii_ew_ecdf_four_snr_bins}
\end{figure*}

{\color{black}To assess whether the \Niii\; EW distribution depends on continuum S/N, we divided the N-loud quasars into four bins based on the continuum S/N measured over the rest-frame 1675--1725~\AA{} window. Figure~\ref{fig:niii_ew_ecdf_four_snr_bins} shows the corresponding cumulative EW distributions. The distributions are broadly similar across the four S/N bins, with median EWs of 4.88, 4.59, 4.59, and 4.40~\AA{} for \(5<S/N\leq7\), \(7<S/N\leq10\), \(10<S/N\leq15\), and \(S/N>15\), respectively. The lowest-S/N bin shows a slight shift toward larger EWs, but the overall variation is small, indicating only a weak dependence of the \Niii\; EW distribution on continuum S/N among the N-loud quasars selected from the continuum-S/N-limited parent sample.}

\begin{figure*}[htbp]
    \centering
    \includegraphics[width=\textwidth]{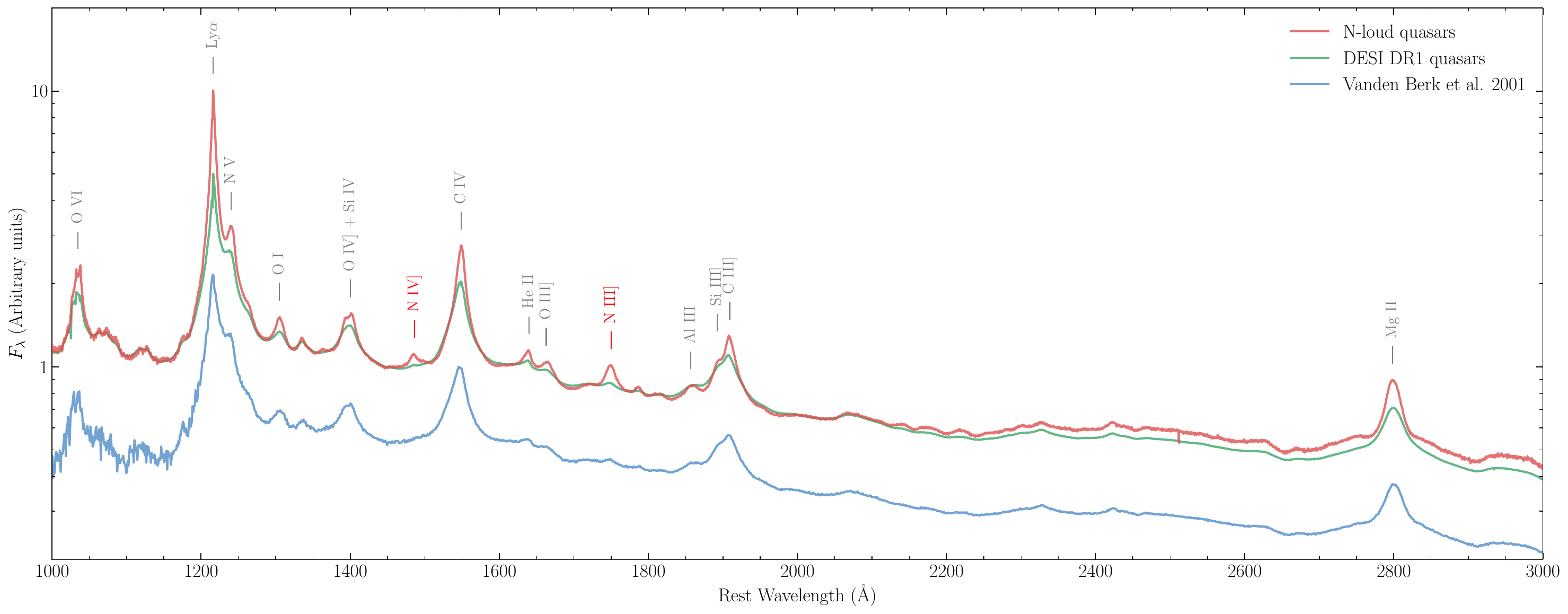}
    \caption{\footnotesize
    {\color{black}Composite spectra for the N-loud quasars (red) and DESI DR1 parent quasars (green).}
    The dashed vertical lines mark the positions of the \Niv\;and \Niii\;lines.
    For comparison, the blue curve denotes the composite spectrum of \citet{Vanden2001}.
    }
    \label{fig:composite_spec}
\end{figure*}

{\color{black} We next extend the comparison beyond the nitrogen lines to the overall UV spectral properties of N-loud quasars.}
Accordingly, we constructed composite spectra for the N-loud quasars and the
DESI DR1 parent sample. All spectra were transformed to the rest frame
and rescaled to a common continuum level at 1450~\AA\ before
combination. We adopted an unweighted average to reduce the influence of
S/N differences among individual spectra. The resulting
composites are shown in Figure~\ref{fig:composite_spec}. The two
composites have broadly similar UV continuum shapes over most of
the wavelength range, but several broad emission features are stronger in
the N-loud composite. The most distinctive differences are the enhanced
broad nitrogen features, including \Nv, \Niv, and \Niii.
{\color{black} These differences confirm that enhanced nitrogen emission is the primary spectral signature of the N-loud population.}

{\color{black} The corresponding emission-line measurements from the composite spectra are summarized in Table~\ref{tab:composite_lines}. Compared with the DESI DR1 composite spectrum, the N-loud composite spectrum shows systematically larger EWs and narrower FWHMs for all of the listed emission lines. Ly$\alpha$ is stronger by about a factor of two, while N V and C IV are enhanced by $\sim$20\% and $\sim$30\%, respectively. The FWHMs are typically smaller by $\sim$10-30\%, with C IV and Mg II narrower by about 25\%.}


{\color{black}
To further examine whether the broad-line differences seen in the composite spectra persist at the individual-object level, we constructed a control sample matched to the DESI N-loud quasars in redshift and UV continuum luminosity, thereby minimizing differences associated with the underlying redshift and luminosity distributions. Because the DESI DR1 quasar catalog does not provide homogeneous measurements of broad-emission-line properties and single-epoch virial black hole masses over the redshift range considered here, we selected the control quasars from the SDSS DR14 quasar spectral catalog of \citet{Rakshit2020}. Details of the control-sample construction are provided in Appendix~\ref{sec:z_L1450_Nloud_sdss}.
}

Figure~\ref{fig:contour_Civ_Mgii} shows that
the N-loud quasars are systematically shifted toward smaller EW and
narrower FWHM in both \Civ\; and \Mgii, with the difference being more
pronounced in FWHM. The median EW offsets are of order \(10~\mathrm{\AA}\), while
the two samples cover broadly similar EW ranges. In contrast, the median
FWHM offsets exceed \(1000~{\rm km~s^{-1}}\), and the N-loud quasars
occupy a substantially narrower FWHM range, particularly for \Mgii. 
A comparison between N-loud quasars and a control sample matched to the
DESI parent-quasar redshift-luminosity distribution also shows the same
qualitative trend (Figure~\ref{fig:contour_Civ_Mgii_desi}; Appendix~\ref{sec:compared_DESI_parent}).
{\color{black}
Although the EW differences vary between the composite spectrum and individual object comparisons, the consistently narrower \Civ\; and \Mgii\;  profiles indicate that reduced line widths in these two lines are a robust spectral property of N-loud quasars.
}


\setcounter{table}{1}
\begin{table*}[htbp]
    \centering
    \caption{{\color{black}Emission-Line Properties of the Composite Spectra}}
    \label{tab:composite_lines}

    \color{black}

    \begin{tabular*}{0.65\textwidth}{@{\extracolsep{\fill}}lcccc@{}}
    \hline\hline
    \multirow{2}{*}{Emission Lines}
        & \multicolumn{2}{c}{N-loud Quasars}
        & \multicolumn{2}{c}{DESI DR1 Quasars} \\
    \cline{2-3}\cline{4-5}
        & EW (\AA)
        & FWHM (km s$^{-1}$)
        & EW (\AA)
        & FWHM (km s$^{-1}$) \\
    \hline
    Ly$\alpha$                        & 72.3 & 2029 & 35.9 & 2810 \\
    \Nv                               & 35.0 & 5601 & 29.2 & 6174 \\
    \ion{O}{1}                        & 3.7  & 3236 & 2.4  & 4032 \\
    \ion{Si}{4} + \ion{O}{4}]         & 10.0 & 4804 & 8.8  & 5923 \\
    \Civ                              & 44.8 & 3416 & 34.5 & 4645 \\
    \ion{O}{3}]                       & 1.9  & 2736 & 0.9  & 2975 \\
    \Ciii                             & 16.9 & 4347 & 14.9 & 5505 \\
    \Mgii                             & 42.1 & 3140 & 36.4 & 4196 \\
    \hline
    \end{tabular*}

    \vskip 2mm

    \begin{minipage}{0.65\textwidth}
        \footnotesize
        \raggedright
        \setlength{\parindent}{0pt}
        \setlength{\parskip}{0pt}
        {\color{black}\textbf{Notes.}
        Measurements are derived from the best-fitting models of the
        N-loud quasar and DESI DR1 quasar composite spectra shown in
        Figure~\ref{fig:composite_spec}.
        Equivalent widths are given in the rest frame, and FWHMs are
        expressed in km s$^{-1}$.}
    \end{minipage}
\end{table*}

\begin{figure*}[htbp]
    \centering
    \includegraphics[width=\textwidth]{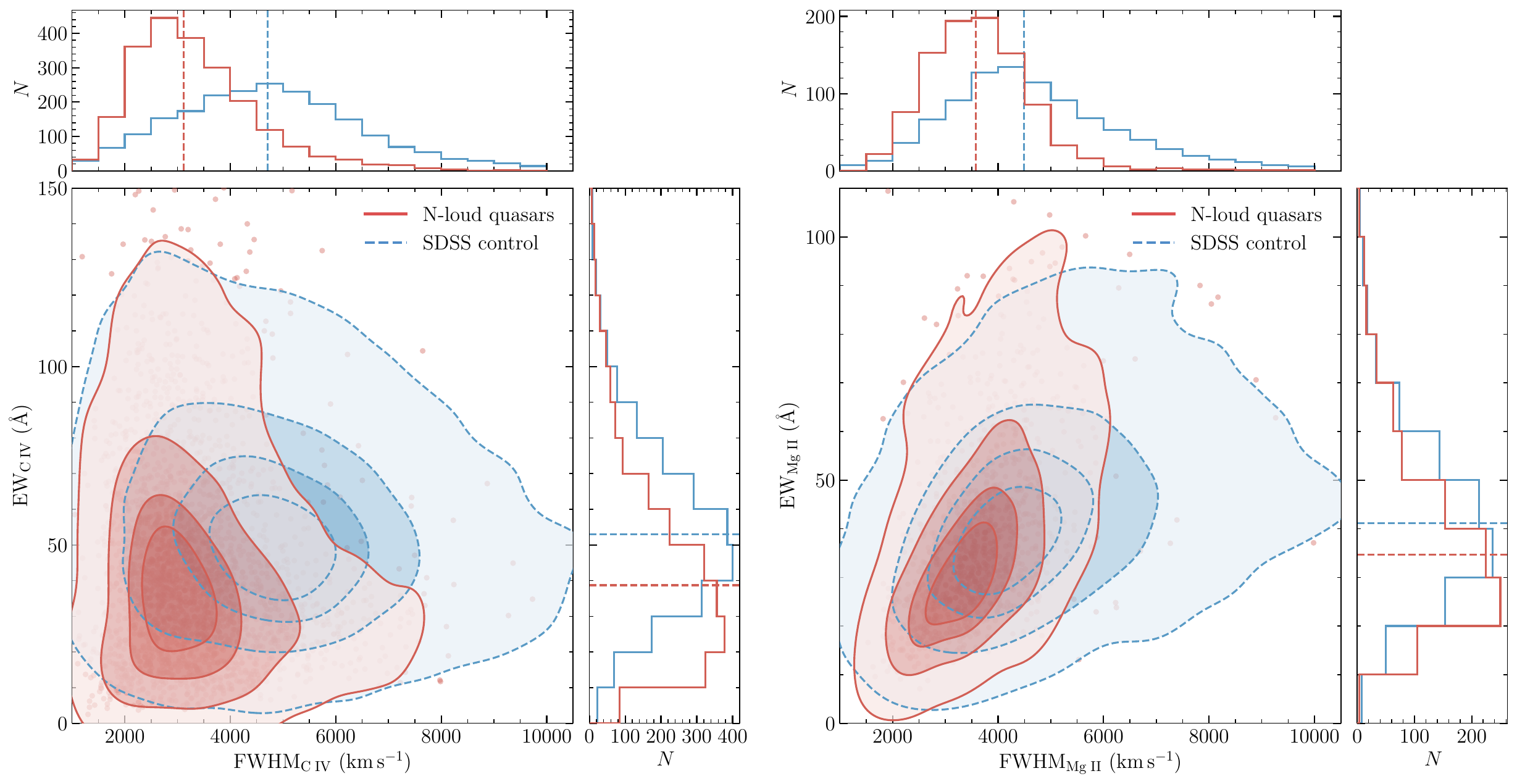}
    \caption{\footnotesize 
    Rest-frame EW versus FWHM distributions of \Civ\;(left panel) and \Mgii\;(right panel) for the N-loud quasars (red solid contours) and SDSS DR14 control sample (blue dashed contours). Red points show individual objects. The 20\%, 40\%, 68\%, and 95\% contours and the corresponding density maps are shown. For each main panel, the top and right panels show the corresponding one-dimensional distributions of FWHM and EW, respectively, with dashed lines indicating the median values.
    }
    \label{fig:contour_Civ_Mgii}
\end{figure*}

\subsection{Black Hole Masses and Eddington Ratios}\label{sec:BHmass}

Motivated by the systematically narrower \Civ\; and \Mgii\; broad lines
of the N-loud quasars, we further estimated their single-epoch virial
black hole masses and Eddington ratios. The black hole masses
(\(M_{\rm BH}\)) were estimated using the broad \Mgii\; and \Civ\;
emission lines. The \Mgii\; estimator was adopted as the fiducial
estimate whenever reliable \Mgii\; measurements were available;
otherwise, the \Civ\; estimator was used. For the \Mgii\; estimator, we
adopted the calibration from \citet{Vestergaard2009},
\begin{equation}
\log \left( \frac{M_{\rm BH}}{M_\odot} \right)
= 0.86 + 0.50 \log \left( \frac{\lambda L_{3000}}{10^{44}\ {\rm erg\ s^{-1}}} \right)
+ 2 \log \left( \frac{{\rm FWHM}_{\rm Mg\,II}}{{\rm km\ s^{-1}}} \right).
\end{equation}
For the \Civ\; estimator, we adopted the calibration from
\citet{Vestergaard2006},
\begin{equation}
\log \left( \frac{M_{\rm BH}}{M_\odot} \right)
= 0.66 + 0.53 \log \left( \frac{\lambda L_{1350}}{10^{44}\ {\rm erg\ s^{-1}}} \right)
+ 2 \log \left( \frac{{\rm FWHM}_{\rm C\,IV}}{{\rm km\ s^{-1}}} \right).
\end{equation}
Here, \(\lambda L_{3000}\) and \(\lambda L_{1350}\) are the
monochromatic continuum luminosities at 3000 and 1350~\AA, respectively,
and \({\rm FWHM}_{\rm Mg\,II}\) and \({\rm FWHM}_{\rm C\,IV}\) are the
full widths at half maximum of the broad \Mgii\; and \Civ\; emission
lines. The continuum luminosities and line widths were taken from our
spectral measurements described in Section~\ref{sec:sample}.

We further calculated the Eddington ratio as
\(\lambda_{\rm Edd}=L_{\rm bol}/L_{\rm Edd}\), where
\(L_{\rm Edd}=1.26\times10^{38}(M_{\rm BH}/M_\odot)\,
{\rm erg\ s^{-1}}\). The bolometric luminosity was estimated from the
monochromatic continuum luminosity at 1350~\AA,
\(L_{\rm bol}=3.81\,L_{1350}\), consistent with \citet{Rakshit2020}.

We note that these single-epoch virial black hole masses are subject to
substantial systematic uncertainties. In particular, \Civ\; may be
affected by non-virial components such as outflows, which can introduce
additional scatter and bias relative to \Mgii-based black hole mass
estimates \citep{Coatman2016,Coatman2017}. Therefore, the black hole
masses and Eddington ratios derived here are used primarily for
statistical characterization of the sample rather than for precise
measurements of individual objects.

\begin{figure*}[htbp]
    \centering
    \includegraphics[width=0.5\textwidth]{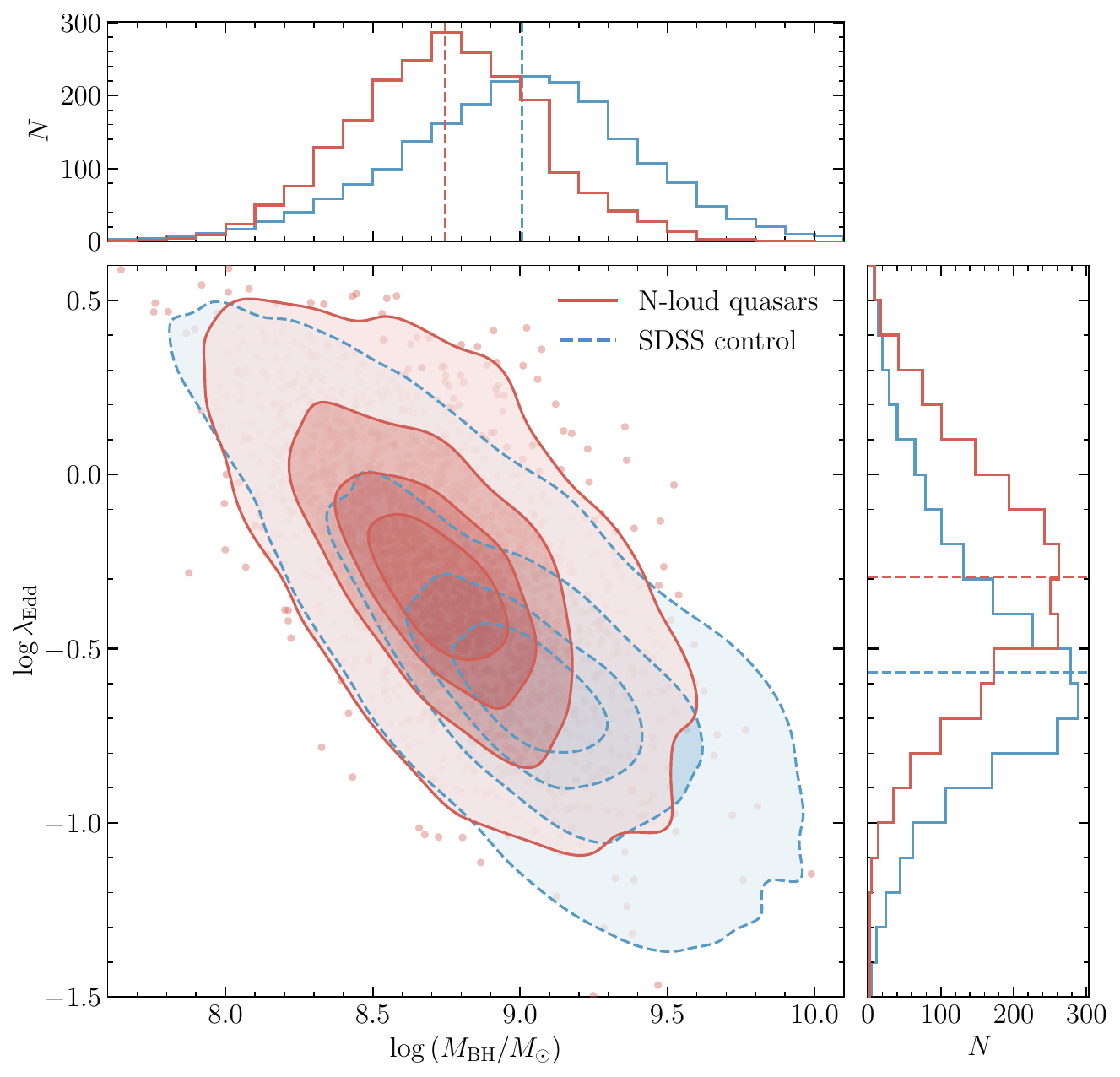}
    \caption{\footnotesize
    Black hole mass versus Eddington ratio distribution for the N-loud quasars and the SDSS control sample. The 20\%, 40\%, 68\%, and 95\% percentile density contours are shown, with shaded regions indicating the corresponding density levels. Red solid contours and points represent the N-loud quasars, while blue dashed contours represent the SDSS DR14 control sample. The top and right panels show the corresponding one-dimensional distributions of $\log M_{\rm BH}$ and $\log \lambda_{\rm Edd}$, respectively, with dashed lines indicating the median values.}
    \label{fig:Mbh_edd}
\end{figure*}

Figure~\ref{fig:Mbh_edd} compares the \(M_{\rm BH}\)--\(\lambda_{\rm Edd}\)
distributions of N-loud quasars and the SDSS control sample constructed
in Appendix~\ref{sec:z_L1450_Nloud_sdss}. Although the two samples
overlap substantially, their median values show that the N-loud quasars
are systematically shifted toward lower black hole masses and higher
Eddington ratios. The median \(\log M_{\rm BH}/M_\odot\) of the N-loud
quasars is about 8.75, approximately 0.25 dex lower than that of the
SDSS control sample, while their median \(\log\lambda_{\rm Edd}\) is
about \(-0.3\), approximately 0.3 dex higher. A comparison between the
N-loud quasars and a control sample matched to the DESI parent-quasar
redshift--luminosity distribution also shows the same qualitative trend
(Figure~\ref{fig:Mbh_edd_desi}; Appendix~\ref{sec:compared_DESI_parent}). The black hole mass offset
is also consistent with the systematically narrower \Civ\; and \Mgii\;
broad lines shown in Figure~\ref{fig:contour_Civ_Mgii}, since
single-epoch virial masses depend quadratically on line width. 
Overall, the N-loud quasars show a statistical trend toward lower black hole masses and higher Eddington ratios in the present study.

{\color{black}
Previous studies have reached broadly consistent conclusions regarding the black hole and accretion properties of N-loud quasars. \citet{Batra2014} found systematically lower virial black hole masses for the most N-loud quasars relative to quasars of similar continuum luminosity. \citet{Matsuoka2017} reported a similar trend using independent \Civ- and H$\beta$-based mass estimates, with N-loud quasars showing lower black hole masses and higher Eddington ratios. In contrast, \citet{Jiang2008} inferred comparable black hole masses for N-loud and normal quasars based on the similar \Mgii\; FWHMs of the two samples. Since their \Mgii\; analysis was restricted to a relatively small, bright subsample at $z<2.1$, whereas the \Mgii\;$+$ \Feii\; fitting procedure was the same as that adopted here, the discrepancy may reflect differences in sample size and sample selection.
}

\subsection{Radio Properties}\label{sec:radio}

{\color{black}We investigated the radio properties of the N-loud sample by cross-matching our catalog with the Very Large Array Sky Survey (VLASS; \citealt{Lacy2020}). VLASS has a resolution of approximately \(2\farcs5\), and we adopted a positional matching radius of \(1\farcs5\).}
A total of {\color{black}217} N-loud quasars have VLASS counterparts under this
criterion. The radio-loudness parameter is defined as
\(R=f_{\rm 6\,cm}/f_{2500}\) \citep{Jiang2007}, where \(f_{\rm 6\,cm}\)
is the rest-frame 6 cm flux density derived from the VLASS radio flux
assuming a power-law radio slope of \(-0.7\), and \(f_{2500}\) is the
flux density at rest-frame 2500~\AA. Quasars with \(R>10\) are classified
as radio-loud.
The VLASS peak fluxes and radio-loudness parameters for the matched sources are listed in Table~\ref{tab:Nloud}.

Among the radio-detected sources, {\color{black}216} satisfy this criterion,
corresponding to a radio-loud fraction of
\({\color{black}\sim9.8\%}\) for the full N-loud sample. The radio-loud
fraction varies among the nitrogen line subtypes: it is
\({\color{black}\sim14.2\%}\) for objects showing both \Niii\; and \Niv,
\({\color{black}\sim9.5\%}\) for those with only \Niii, and
\({\color{black}\sim6.6\%}\) for those with only \Niv. These subtype-dependent fractions
should be interpreted with caution, especially for the \Niv-only class,
which contains only a small number of objects. {\color{black}The radio-loud fraction derived here is substantially lower than that reported by \citet{Jiang2008}. Differences in parent-sample selection should be considered when comparing these fractions. The SDSS DR5 quasar catalog used by \citet{Jiang2008} included targeting based on the Faint Images of the Radio Sky at Twenty-cm survey (FIRST; \citealt{Becker1995}), whereas VLASS information in this work was introduced only through cross-matching after the N-loud sample had been defined. This selection difference may contribute to the observed difference in radio-loud fraction. The fraction of quasars targeted on the basis of FIRST counterparts in their parent sample was not reported by \citet{Jiang2008}, leaving the magnitude of this contribution uncertain.} Overall, these results show that a non-negligible
fraction of DESI N-loud quasars are radio-loud, with the highest
radio-loud fraction found among objects showing both \Niii\; and \Niv\;
emission.

\setcounter{table}{0}
\begin{sidewaystable}
\centering
\caption{DESI DR1 N-loud quasar catalog}
\label{tab:Nloud}
\color{black}
\setlength{\tabcolsep}{3pt}
\renewcommand{\arraystretch}{0.92}
\scriptsize
\begin{tabular}{lcccccccccccccccc}
\hline\hline
R.A. & Decl. & $z$ & $\log L_{1450}$ & $\alpha$ & 
\multicolumn{4}{c}{EW (\AA)} & 
\multicolumn{4}{c}{FWHM (km s$^{-1}$)} & 
\Feii/\Mgii & Peak Flux & $R$ \\
\cline{6-9}\cline{10-13}
&  &  & (erg s$^{-1}$) & 
& \Niv & \Niii & \Civ & \Mgii
& \Niv & \Niii & \Civ & \Mgii
&  & (mJy\,beam$^{-1}$) & \\
(1) & (2) & (3) & (4) & (5)
& (6) & (7) & (8) & (9) & (10)
& (11) & (12) & (13) & (14) & (15)
& (16) \\
\hline
0.12 & 17.45 & 2.60 & $46.28\pm0.01$ & $-1.54\pm0.01$ & $\ldots$ & $3.3\pm0.3$ & $42.8\pm1.5$ & $\ldots$ & $\ldots$ & $2955\pm307$ & $3644\pm284$ & $\ldots$ & $\ldots$ & $\ldots$ & $\ldots$ \\
0.14 & 3.66 & 3.22 & $46.32\pm0.01$ & $-1.22\pm0.02$ & $3.8\pm0.5$ & $3.7\pm0.4$ & $63.4\pm1.3$ & $\ldots$ & $5926\pm763$ & $4593\pm486$ & $3681\pm347$ & $\ldots$ & $\ldots$ & $\ldots$ & $\ldots$ \\
0.45 & 4.89 & 2.13 & $45.79\pm0.01$ & $-1.33\pm0.06$ & $3.6\pm0.9$ & $\ldots$ & $93.2\pm1.4$ & $47.8\pm1.6$ & $3004\pm558$ & $\ldots$ & $1913\pm118$ & $3491\pm230$ & $3.0\pm0.2$ & $\ldots$ & $\ldots$ \\
0.70 & -10.06 & 2.52 & $46.09\pm0.01$ & $-1.15\pm0.03$ & $\ldots$ & $6.3\pm0.4$ & $26.6\pm1.2$ & $\ldots$ & $\ldots$ & $3169\pm213$ & $2646\pm455$ & $\ldots$ & $\ldots$ & $\ldots$ & $\ldots$ \\
0.86 & 5.98 & 3.76 & $46.22\pm0.01$ & $-1.84\pm0.05$ & $4.9\pm0.8$ & $3.6\pm0.6$ & $75.9\pm0.8$ & $\ldots$ & $4355\pm630$ & $2307\pm285$ & $2284\pm70$ & $\ldots$ & $\ldots$ & $\ldots$ & $\ldots$ \\
0.98 & -0.04 & 2.49 & $45.89\pm0.01$ & $-1.30\pm0.03$ & $5.0\pm0.5$ & $4.4\pm0.3$ & $17.4\pm0.5$ & $\ldots$ & $4550\pm462$ & $1910\pm148$ & $3881\pm113$ & $\ldots$ & $\ldots$ & $\ldots$ & $\ldots$ \\
1.43 & 5.86 & 2.03 & $45.80\pm0.02$ & $-1.66\pm0.04$ & $3.7\pm0.5$ & $5.4\pm0.4$ & $25.1\pm1.5$ & $17.4\pm0.7$ & $2052\pm284$ & $1913\pm173$ & $2181\pm561$ & $2733\pm110$ & $2.9\pm0.3$ & $\ldots$ & $\ldots$ \\
1.45 & 19.89 & 3.17 & $45.92\pm0.01$ & $-1.85\pm0.03$ & $\ldots$ & $3.5\pm0.4$ & $2.5\pm0.3$ & $\ldots$ & $\ldots$ & $1385\pm187$ & $1303\pm192$ & $\ldots$ & $\ldots$ & $99.3\pm0.1$ & 12795 \\
1.62 & 16.26 & 1.97 & $45.66\pm0.01$ & $-1.83\pm0.04$ & $4.8\pm0.8$ & $11.3\pm0.8$ & $19.3\pm1.6$ & $19.5\pm0.9$ & $3254\pm580$ & $4594\pm331$ & $1556\pm282$ & $3085\pm142$ & $2.3\pm0.3$ & $\ldots$ & $\ldots$ \\
1.64 & 1.66 & 2.81 & $45.99\pm0.02$ & $-0.49\pm0.03$ & $\ldots$ & $5.5\pm0.6$ & $15.4\pm0.9$ & $\ldots$ & $\ldots$ & $4153\pm324$ & $4153\pm227$ & $\ldots$ & $\ldots$ & $3.2\pm0.2$ & 103 \\
1.70 & -9.68 & 1.83 & $45.65\pm0.02$ & $-0.68\pm0.05$ & $\ldots$ & $4.0\pm0.5$ & $67.6\pm2.6$ & $53.4\pm1.6$ & $\ldots$ & $2652\pm350$ & $3490\pm281$ & $4300\pm65$ & $3.5\pm0.1$ & $\ldots$ & $\ldots$ \\
1.79 & 15.73 & 1.99 & $45.90\pm0.02$ & $-1.21\pm0.02$ & $\ldots$ & $4.2\pm0.3$ & $67.8\pm1.5$ & $38.0\pm1.0$ & $\ldots$ & $2048\pm169$ & $2797\pm213$ & $3267\pm48$ & $4.1\pm0.1$ & $\ldots$ & $\ldots$ \\
2.00 & 15.14 & 1.97 & $46.03\pm0.01$ & $-1.37\pm0.02$ & $3.9\pm0.7$ & $2.8\pm0.4$ & $62.8\pm1.8$ & $27.4\pm0.8$ & $5794\pm1137$ & $2667\pm386$ & $2297\pm202$ & $2832\pm52$ & $4.2\pm0.2$ & $\ldots$ & $\ldots$ \\
2.30 & 0.97 & 2.70 & $46.02\pm0.01$ & $-1.55\pm0.03$ & $6.5\pm0.8$ & $2.0\pm0.3$ & $34.3\pm0.6$ & $\ldots$ & $8405\pm1098$ & $1858\pm285$ & $4875\pm92$ & $\ldots$ & $\ldots$ & $\ldots$ & $\ldots$ \\
2.92 & 0.90 & 2.36 & $45.86\pm0.01$ & $-1.50\pm0.02$ & $3.2\pm0.6$ & $2.5\pm0.3$ & $45.0\pm0.8$ & $35.2\pm1.5$ & $6211\pm1136$ & $1786\pm240$ & $2110\pm182$ & $3712\pm97$ & $2.2\pm0.2$ & $\ldots$ & $\ldots$ \\
3.01 & 2.31 & 3.53 & $46.41\pm0.01$ & $-1.29\pm0.01$ & $\ldots$ & $4.7\pm0.2$ & $24.6\pm0.2$ & $\ldots$ & $\ldots$ & $3279\pm186$ & $2569\pm24$ & $\ldots$ & $\ldots$ & $\ldots$ & $\ldots$ \\
3.22 & 11.25 & 1.67 & $46.08\pm0.01$ & $-1.72\pm0.05$ & $3.6\pm0.6$ & $\ldots$ & $121.2\pm1.3$ & $94.9\pm1.0$ & $3473\pm478$ & $\ldots$ & $4959\pm175$ & $4632\pm79$ & $2.3\pm0.1$ & $\ldots$ & $\ldots$ \\
3.25 & -12.71 & 2.75 & $46.79\pm0.02$ & $-1.06\pm0.01$ & $\ldots$ & $4.9\pm0.2$ & $33.0\pm0.5$ & $\ldots$ & $\ldots$ & $6988\pm337$ & $2888\pm120$ & $\ldots$ & $\ldots$ & $\ldots$ & $\ldots$ \\
3.25 & -2.41 & 3.30 & $46.02\pm0.01$ & $-1.32\pm0.03$ & $\ldots$ & $5.1\pm0.6$ & $37.6\pm1.1$ & $\ldots$ & $\ldots$ & $3741\pm506$ & $2115\pm245$ & $\ldots$ & $\ldots$ & $\ldots$ & $\ldots$ \\
3.37 & 1.11 & 3.07 & $46.62\pm0.01$ & $-1.57\pm0.01$ & $\ldots$ & $3.6\pm0.1$ & $11.6\pm0.4$ & $\ldots$ & $\ldots$ & $6289\pm915$ & $4306\pm91$ & $\ldots$ & $\ldots$ & $\ldots$ & $\ldots$ \\
3.69 & 0.69 & 3.61 & $45.91\pm0.01$ & $-1.48\pm0.04$ & $3.8\pm0.7$ & $\ldots$ & $84.9\pm1.2$ & $\ldots$ & $3205\pm443$ & $\ldots$ & $3724\pm118$ & $\ldots$ & $\ldots$ & $\ldots$ & $\ldots$ \\
3.75 & 20.97 & 3.10 & $46.09\pm0.01$ & $-1.53\pm0.02$ & $\ldots$ & $5.4\pm0.6$ & $56.4\pm1.5$ & $\ldots$ & $\ldots$ & $5258\pm693$ & $2820\pm344$ & $\ldots$ & $\ldots$ & $\ldots$ & $\ldots$ \\
3.84 & 2.83 & 2.35 & $45.92\pm0.01$ & $-1.53\pm0.03$ & $5.6\pm0.5$ & $\ldots$ & $88.5\pm1.0$ & $60.3\pm1.6$ & $3816\pm272$ & $\ldots$ & $2150\pm69$ & $3491\pm194$ & $2.7\pm0.1$ & $\ldots$ & $\ldots$ \\
4.03 & 0.59 & 3.07 & $45.74\pm0.02$ & $-1.19\pm0.04$ & $\ldots$ & $3.9\pm0.5$ & $78.7\pm2.1$ & $\ldots$ & $\ldots$ & $2782\pm395$ & $2562\pm180$ & $\ldots$ & $\ldots$ & $\ldots$ & $\ldots$ \\
4.26 & 22.91 & 2.84 & $46.06\pm0.01$ & $-1.51\pm0.02$ & $\ldots$ & $3.3\pm0.3$ & $28.8\pm0.6$ & $\ldots$ & $\ldots$ & $2807\pm301$ & $1879\pm35$ & $\ldots$ & $\ldots$ & $\ldots$ & $\ldots$ \\
4.40 & -11.88 & 2.61 & $45.91\pm0.01$ & $-0.66\pm0.03$ & $\ldots$ & $3.1\pm0.4$ & $55.9\pm2.9$ & $\ldots$ & $\ldots$ & $2038\pm296$ & $3207\pm362$ & $\ldots$ & $\ldots$ & $\ldots$ & $\ldots$ \\
4.53 & 4.26 & 2.00 & $45.72\pm0.02$ & $-1.85\pm0.04$ & $\ldots$ & $3.8\pm0.6$ & $25.9\pm1.2$ & $31.2\pm1.2$ & $\ldots$ & $2683\pm463$ & $6455\pm282$ & $3497\pm130$ & $4.5\pm0.2$ & $\ldots$ & $\ldots$ \\
4.60 & 19.62 & 1.95 & $46.01\pm0.01$ & $-1.32\pm0.02$ & $\ldots$ & $6.2\pm0.6$ & $17.9\pm0.8$ & $19.3\pm0.5$ & $\ldots$ & $7780\pm773$ & $4500\pm298$ & $2592\pm52$ & $3.7\pm0.3$ & $\ldots$ & $\ldots$ \\
4.86 & 10.53 & 2.11 & $46.17\pm0.01$ & $-1.00\pm0.02$ & $\ldots$ & $6.8\pm0.4$ & $51.7\pm1.2$ & $32.1\pm0.7$ & $\ldots$ & $5055\pm449$ & $4070\pm233$ & $3920\pm55$ & $2.8\pm0.1$ & $\ldots$ & $\ldots$ \\
5.03 & 22.68 & 2.50 & $46.05\pm0.01$ & $-1.34\pm0.03$ & $\ldots$ & $4.2\pm0.4$ & $9.2\pm0.7$ & $\ldots$ & $\ldots$ & $2626\pm307$ & $3871\pm295$ & $\ldots$ & $\ldots$ & $2.3\pm0.1$ & 99 \\
5.15 & 17.85 & 3.10 & $46.14\pm0.01$ & $-1.50\pm0.02$ & $\ldots$ & $5.0\pm0.5$ & $28.9\pm1.6$ & $\ldots$ & $\ldots$ & $4277\pm436$ & $3126\pm298$ & $\ldots$ & $\ldots$ & $\ldots$ & $\ldots$ \\
5.62 & 23.16 & 1.75 & $45.67\pm0.02$ & $-1.54\pm0.04$ & $\ldots$ & $4.6\pm0.5$ & $62.1\pm3.7$ & $51.6\pm2.3$ & $\ldots$ & $1698\pm224$ & $2934\pm380$ & $2991\pm69$ & $2.9\pm0.2$ & $\ldots$ & $\ldots$ \\
5.69 & 6.39 & 2.14 & $45.84\pm0.01$ & $-1.30\pm0.02$ & $\ldots$ & $3.2\pm0.4$ & $28.3\pm1.7$ & $30.2\pm1.0$ & $\ldots$ & $3055\pm376$ & $3422\pm338$ & $2921\pm58$ & $4.7\pm0.2$ & $\ldots$ & $\ldots$ \\
\hline
\end{tabular}
\vskip 2mm
\begin{minipage}{0.94\textheight} 
\scriptsize
\raggedright
\setlength{\parindent}{0pt}
\setlength{\parskip}{0pt}
\setlength{\baselineskip}{0.92\baselineskip}
\textbf{Notes.}
Column (1): R.A. Column (2): Decl. Column (3): redshift.
Column (4): $\log L_{1450}$. Column (5): $\alpha$.
Columns (6)--(9): EW of \Niv, \Niii, \Civ, and \Mgii.
Columns (10)--(13): FWHM of \Niv, \Niii, \Civ, and \Mgii.
Column (14): \Feii/\Mgii\, flux ratio, measured from the fitted \Feii\; UV pseudo-continuum and \Mgii\, emission line.
Column (15): peak flux at 3\,GHz (mJy\,beam$^{-1}$).
Column (16): radio loudness $R$.
Table~\ref{tab:Nloud} is published in its entirety in the electronic edition. A portion is shown here for guidance regarding its form and content.
\end{minipage}
\end{sidewaystable}

\section{Summary}\label{sec:sum}

We present a sample of {\color{black}2,198} DESI DR1 quasars with strong broad
\Niii\; and/or \Niv\; emission lines at \(1.6<z<4.3\), corresponding to
about {\color{black}1.3\%} of the parent quasar sample. The following points summarize
our main findings:

\begin{enumerate}
\item The N-loud quasars have a \(L_{1450}\) distribution broadly similar
to that of the DESI parent sample, but their redshift distribution
differs from that of the parent sample. They are more concentrated
around \(z\sim2.5\)--3, and the {\color{black} corresponding N-loud fraction} reaches a peak of about 2\% near this redshift range.

\item The composite spectrum of the N-loud quasars has a broadly similar
UV continuum shape to that of the parent quasars, while showing
enhanced broad nitrogen emission lines, including \Nv, \Niv, and \Niii.
Other metal emission features also show a moderate enhancement.

\item Relative to a control sample matched in redshift and UV continuum
luminosity, the N-loud quasars systematically show narrower broad
\Civ\; and \Mgii\; emission lines, lower single-epoch virial black hole
masses, and higher Eddington ratios. These tendencies are also seen in
comparison with the parent quasar sample. Relative to the matched control
sample, the median \(\log M_{\rm BH}\) of the N-loud quasars is lower by
about 0.25 dex, while the median \(\log\lambda_{\rm Edd}\) is higher by
about 0.3 dex.

\item The radio-loud fraction of the N-loud quasars is {\color{black}9.8\%}, with the
highest fraction found in objects showing both \Niii\; and \Niv\;
emission.
\end{enumerate}

Overall, these results indicate that N-loud quasars do not simply
represent a small number of extreme objects in the distribution of
nitrogen line strength, but instead show a set of correlated systematic
properties, including a non-uniform redshift distribution, narrower broad
emission lines, lower single-epoch virial black hole masses, higher
Eddington ratios, and a non-negligible radio-loud subgroup. These
properties suggest that N-loud quasars may preferentially appear during a
relatively rapid black hole accretion phase. The consistent enhancement
of multiple broad nitrogen emission lines also suggests that nitrogen
enhancement is part of the overall UV spectral properties of
these objects, rather than an isolated anomaly in a single line. The DESI
N-loud quasar catalog presented here provides a large statistical
baseline for future work. In the next paper of this series, we will use
multiple UV emission lines and photoionization models to analyze
their abundance patterns, testing whether the strong nitrogen emission
reflects an overall increase in metallicity or selective nitrogen
enhancement relative to other elements, and thereby further constraining
the chemical-enrichment mechanism of N-loud quasars and its connection
with black hole accretion state.


\begin{acknowledgments}

S.Z. is grateful to C. Hu and P. Du for valuable discussions. This study is supported by the National Natural Science Foundation of China (NSFC) under grant No. 12588202, the Strategic Priority Research Program of the Chinese Academy of Sciences under grant No. XDB1160103, the National Key R\&D Program of China under grant Nos. 2024YFA1611903, 2023YFE0107800, and 2024YFA1611601, and the CAS Project for Young Scientists in Basic Research under grant No. YSBR-092.

{\color{black}
This research used data obtained with the DESI. DESI construction and operations is managed by the Lawrence Berkeley National Laboratory. This material is based upon work supported by the \href{https://www.energy.gov/}{U.S. Department of Energy}, Office of Science, Office of High-Energy Physics, under Contract No. DE--AC02--05CH11231, and by the National Energy Research Scientific Computing Center, a DOE Office of Science User Facility under the same contract. Additional support for DESI was provided by the \href{https://www.nsf.gov/}{U.S. National Science Foundation} (NSF), Division of Astronomical Sciences under Contract No. AST-0950945 to the NSF's National Optical-Infrared Astronomy Research Laboratory; the \href{https://stfc.ukri.org/}{Science and Technology Facilities Council of the United Kingdom}; the \href{https://www.moore.org/}{Gordon and Betty Moore Foundation}; the \href{https://www.hsfoundation.org/}{Heising-Simons Foundation}; the \href{https://www.cea.fr/}{French Alternative Energies and Atomic Energy Commission} (CEA); the \href{https://secihti.mx/}{National Council of Humanities, Science and Technology of Mexico} (CONAHCYT); the \href{http://www.mineco.gob.es/}{Ministry of Science and Innovation of Spain} (MICINN), and by the DESI Member Institutions: \href{https://www.desi.lbl.gov/collaborating-institutions}{www.desi.lbl.gov/collaborating-institutions}. The DESI collaboration is honored to be permitted to conduct scientific research on I'oligam Du'ag (Kitt Peak), a mountain with particular significance to the \href{https://www.tonation-nsn.gov/}{Tohono O'odham Nation}. Any opinions, findings, and conclusions or recommendations expressed in this material are those of the author(s) and do not necessarily reflect the views of the U.S. National Science Foundation, the U.S. Department of Energy, or any of the listed funding agencies.
}

\end{acknowledgments}



\appendix

\section{Robustness of the Redshift Dependence of the N-loud Fraction to Continuum S/N Selection} \label{sec:SN_redshift}

\begin{figure*}[htbp]
    \centering
    \includegraphics[width=0.5\textwidth]{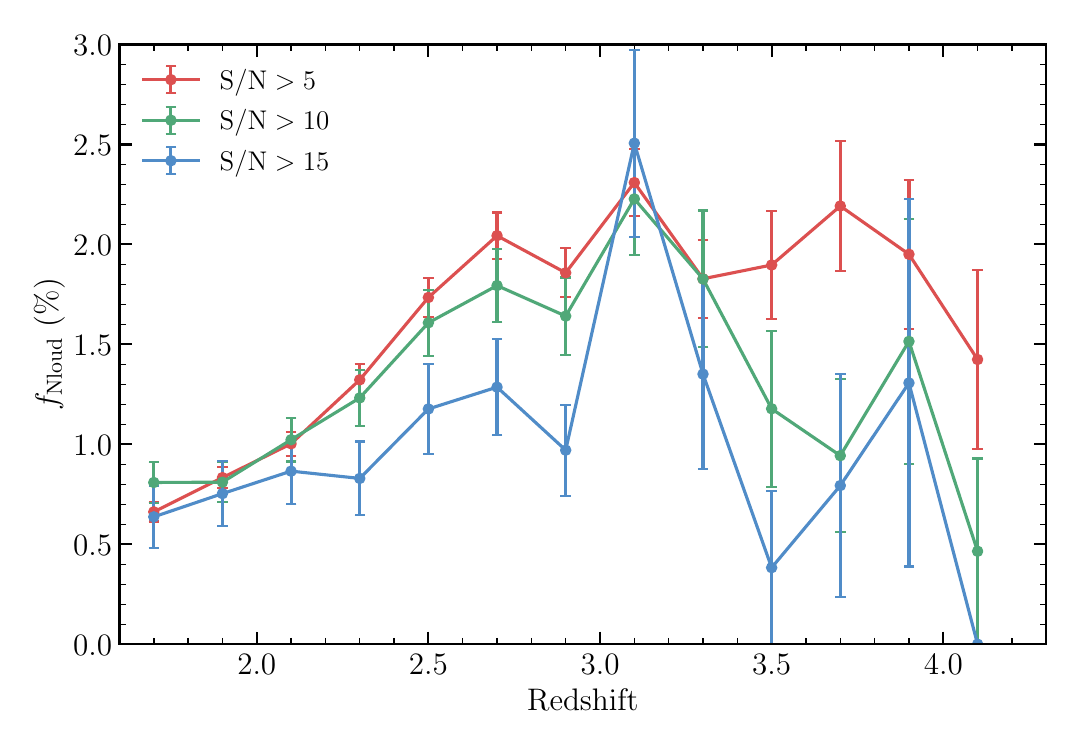}
    \caption{\footnotesize
    Redshift evolution of the N-loud fraction under different continuum S/N selections.
    The red, green, and blue points show the N-loud fractions measured with continuum S/N thresholds of S/N \(>5\), S/N \(>10\), and S/N \(>15\), respectively.
    }
    \label{fig:SNR_robustness}
\end{figure*}

To test whether the observed {\color{black} redshift dependence} of the N-loud fraction is affected by the continuum spectral quality of the parent quasar sample, we remeasured the N-loud fraction using different continuum S/N cuts, including S/N \(>5\), S/N \(>10\), and S/N \(>15\). In each redshift bin, we counted the numbers of parent quasars and N-loud quasars satisfying the corresponding continuum S/N criterion, and then calculated the N-loud fraction, with uncertainties estimated using the binomial approximation. Figure~\ref{fig:SNR_robustness} shows the redshift dependence of the N-loud fraction obtained under different continuum S/N selections. The three curves show broadly consistent {\color{black} redshift-dependent behavior}: the N-loud fraction increases from \(z\sim1.6\) to \(z\sim3\), reaches a peak around \(z\sim3\), and then declines toward higher redshift. Stricter continuum S/N cuts lead to larger uncertainties, especially at the high-redshift end, where the number of quasars with high continuum S/N becomes limited. 
{\color{black} Nevertheless, the overall redshift-dependent trend remains clearly visible under all three selections, indicating that the observed redshift dependence of the N-loud fraction is not primarily driven by the specific continuum S/N threshold adopted here.}

%

\section{Construction of the SDSS Control Sample Matched to DESI N-loud Quasars} \label{sec:z_L1450_Nloud_sdss}

\begin{figure*}[htbp]
    \centering
    \includegraphics[width=0.5\textwidth]{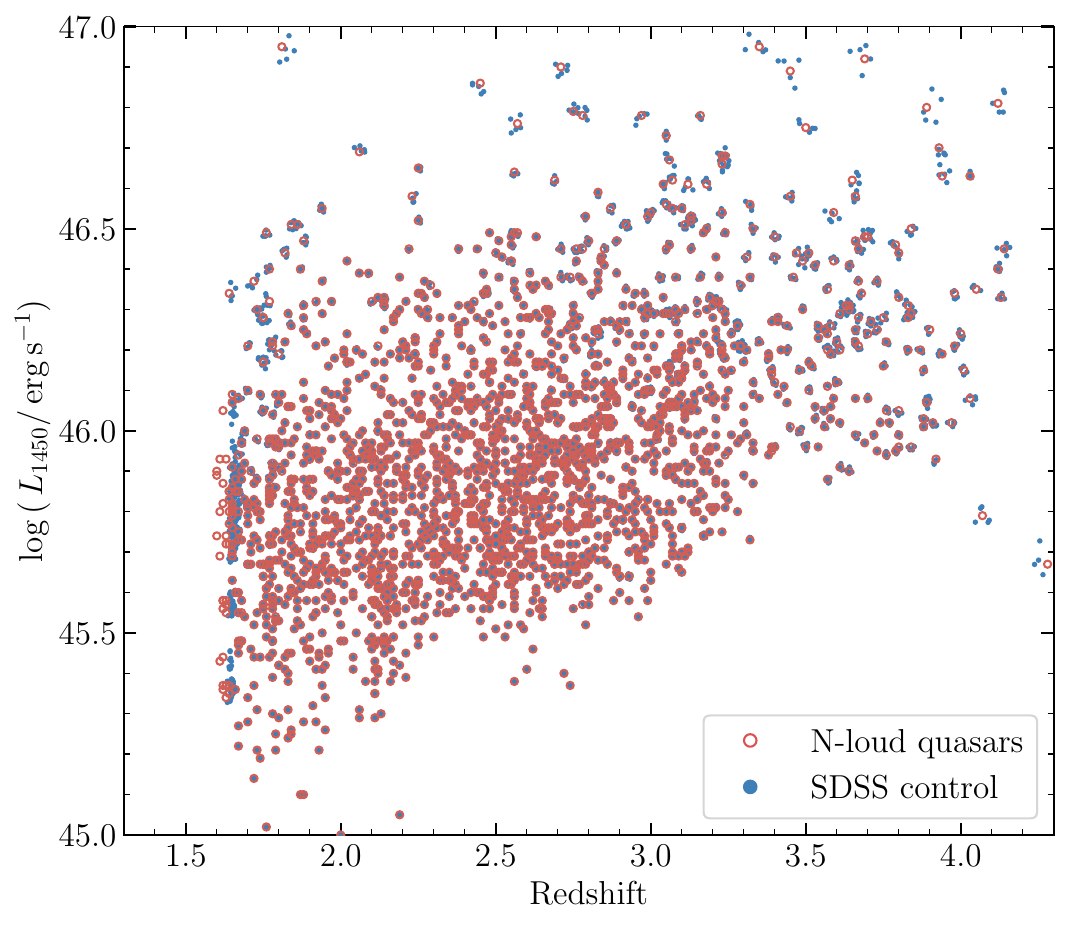}
    \caption{\footnotesize
    Redshift versus rest-frame 1450~\AA{} luminosity distribution for the DESI N-loud quasars and the matched SDSS control sample.
    Red open circles denote the DESI N-loud quasars, and blue filled circles denote the SDSS control quasars selected from the SDSS DR14 quasar catalog.
    }
    \label{fig:contour_z_L1450_Nloud_sdss}
\end{figure*}

To construct a direct comparison sample for the DESI N-loud quasars, we selected SDSS control quasars matched to the DESI N-loud sample in redshift and UV continuum luminosity. We retained SDSS DR14 quasars with reliable continuum luminosity measurements by requiring \texttt{QUALITY\_L1350 = 0}, and converted \(L_{1350}\) to \(L_{1450}\) assuming \(f_{\lambda}\propto \lambda^{-1.5}\). To avoid duplicated sources, SDSS quasars within 2 arcsec of any DESI N-loud quasar were removed through sky-coordinate matching. For each DESI N-loud quasar, we then selected five nearest SDSS quasars in the redshift--\(\log L_{1450}\) plane, without replacement. Figure~\ref{fig:contour_z_L1450_Nloud_sdss} shows that the resulting SDSS control sample closely matches the DESI N-loud sample in both redshift and UV continuum luminosity, providing a suitable control sample for direct comparison with the DESI N-loud quasars in subsequent analyses.

\section{N-loud Quasars Compared with a DESI-parent-like SDSS Control Sample} \label{sec:compared_DESI_parent}

\begin{figure*}[htbp]
    \centering
    \includegraphics[width=0.5\textwidth]{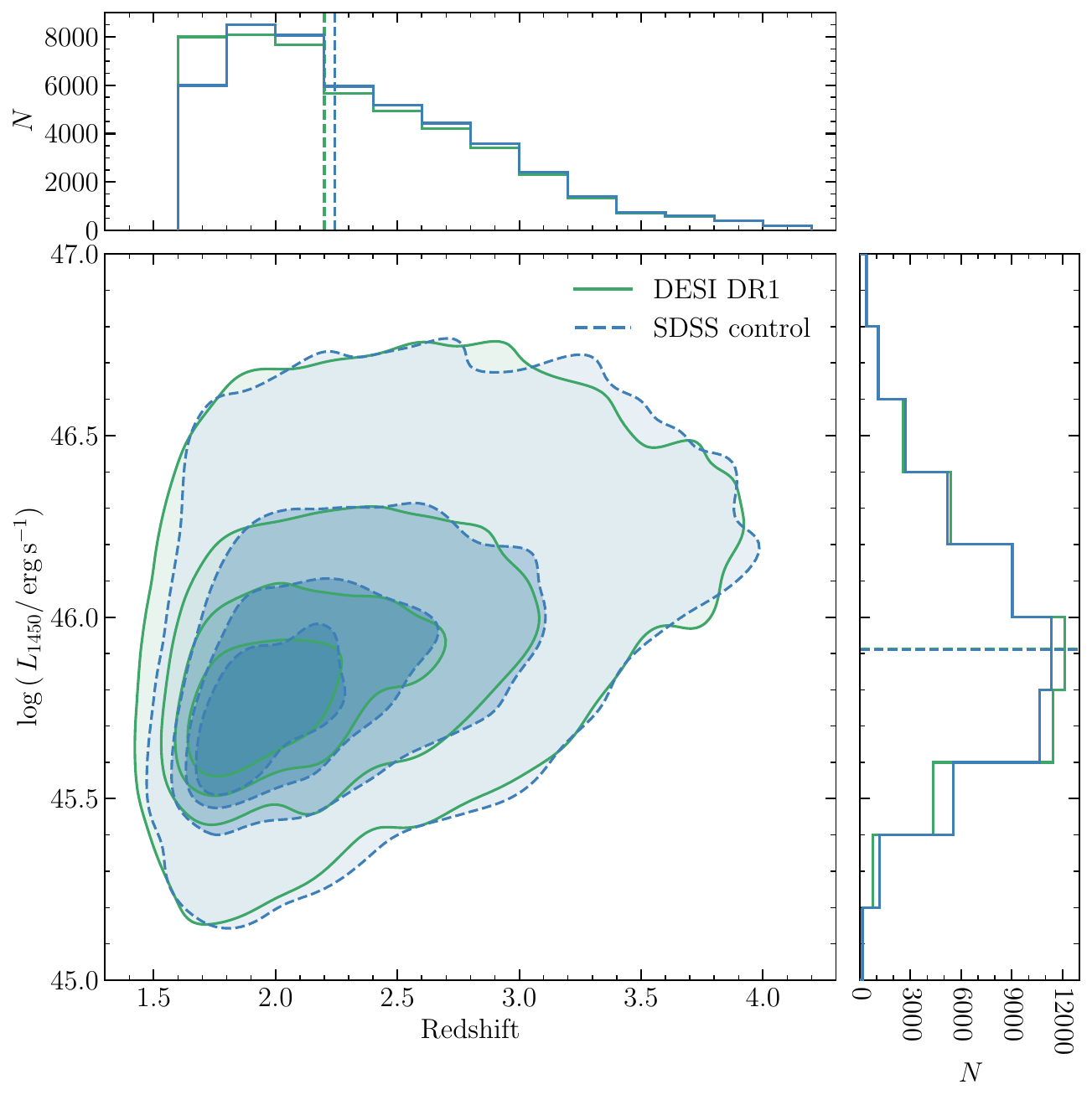}
    \caption{\footnotesize
    Redshift versus rest-frame 1450~\AA{} luminosity distribution for the DESI DR1 parent quasar sample and the SDSS control sample. The 20\%, 40\%, 68\%, and 95\% percentile density contours are shown, with shaded regions indicating the corresponding density levels. Green solid contours represent the DESI DR1 parent quasar sample, while blue dashed contours represent the SDSS DR14 control sample. The top and right panels show the corresponding one-dimensional distributions of redshift and $\log L_{1450}$, respectively, with dashed lines indicating the median values.}
    \label{fig:contour_z_L1450_desi_sdss}
\end{figure*}

To examine the statistical properties of DESI N-loud quasars at the parent-population level, we constructed an SDSS control sample with the same redshift and \(1450\,\text{\AA}\) continuum-luminosity distributions as the DESI parent quasar sample, and compared the N-loud quasars with this control sample. We selected sources from the SDSS DR14 quasar spectral catalog with reliable bolometric luminosities and reddening corrections, requiring \texttt{QUALITY\_LBOL = 0} and \texttt{QUALITY\_REDD = 0}, and converted the SDSS catalog \(\log L_{1350}\) measurements to \(\log L_{1450}\) by assuming \(f_{\lambda}\propto\lambda^{-1.5}\). We then constructed two-dimensional bins over \(1.6<z<4.3\) and \(44.5<\log L_{1450}<47.5\), with bin sizes of \(\Delta z=0.2\) and \(\Delta\log L_{1450}=0.2\). In each bin, SDSS quasars were randomly sampled without replacement according to the fractional occupancy of the DESI parent sample; the target total size was set to 50,000, with the number of sampled objects in each bin determined jointly by this fraction and the number of available SDSS sources in that bin. The resulting SDSS control sample closely matches the DESI parent sample in the overall redshift and \(\log L_{1450}\) distributions, as shown in Figure~\ref{fig:contour_z_L1450_desi_sdss}.

\begin{figure*}[htbp]
    \centering
    \includegraphics[width=\textwidth]{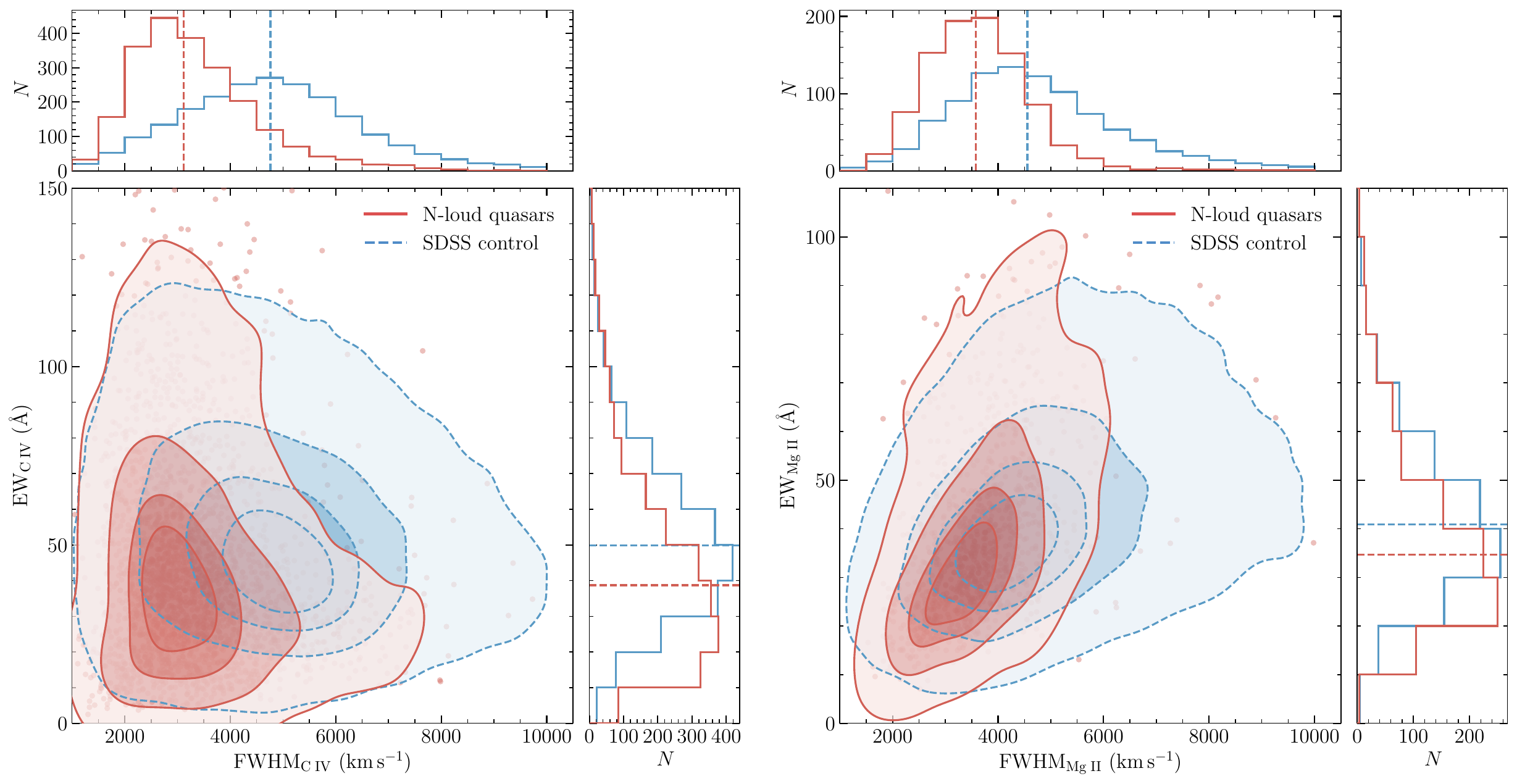}
    \caption{\footnotesize 
    Rest-frame EW versus FWHM distributions of \Civ\;(left panel) and \Mgii\;(right panel) for the N-loud quasars (red solid contours) and SDSS DR14 control sample (blue dashed contours). Red points show individual objects. The 20\%, 40\%, 68\%, and 95\% contours and the corresponding density maps are shown. For each main panel, the top and right panels show the corresponding one-dimensional distributions of FWHM and EW, respectively, with dashed lines indicating the median values.
    }
    \label{fig:contour_Civ_Mgii_desi}
\end{figure*}

\begin{figure*}[htbp]
    \centering
    \includegraphics[width=0.5\textwidth]{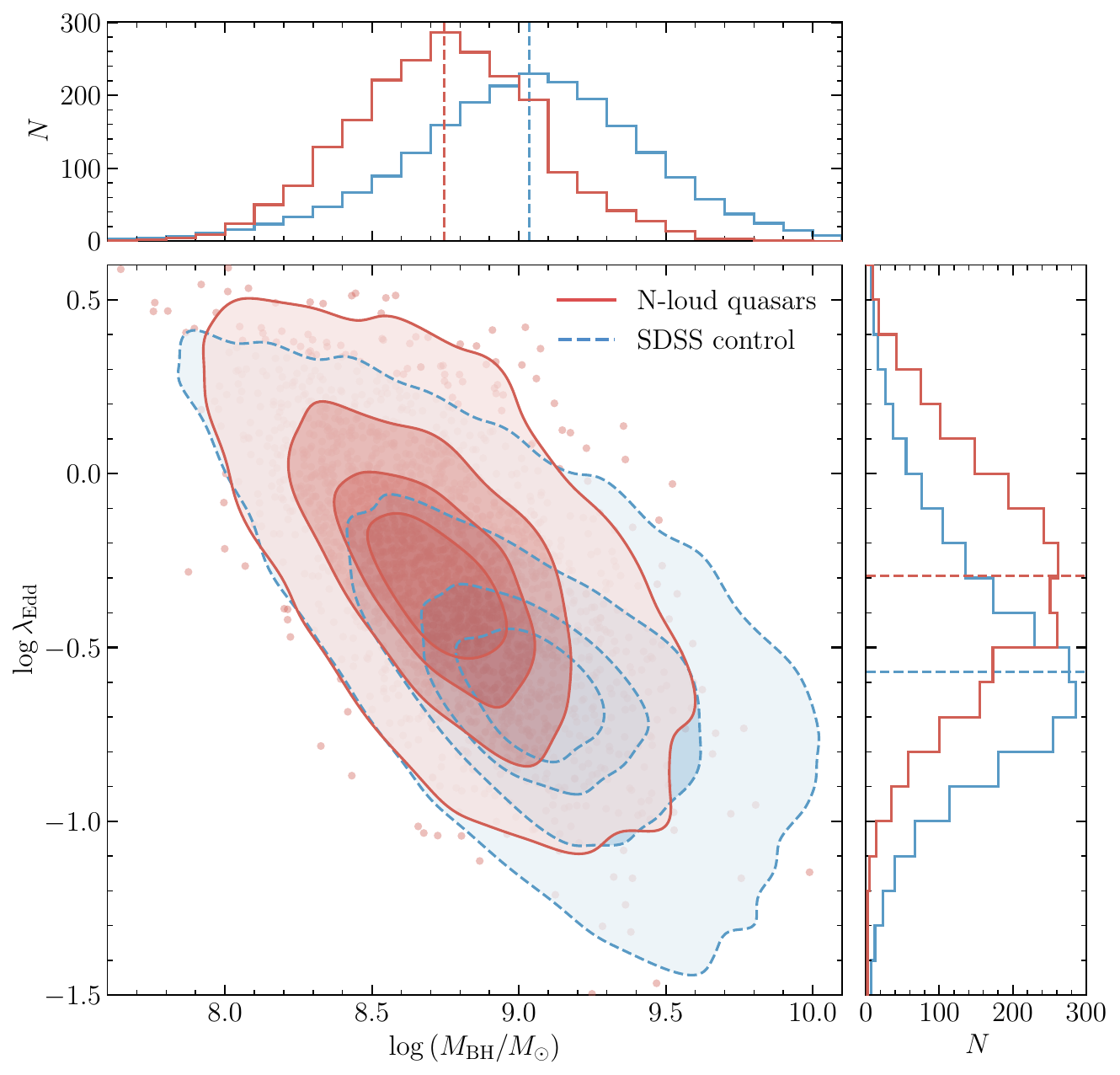}
    \caption{\footnotesize
    Black hole mass versus Eddington ratio distribution for the N-loud quasars and the SDSS control sample. The 20\%, 40\%, 68\%, and 95\% percentile density contours are shown, with shaded regions indicating the corresponding density levels. Red solid contours and points represent the N-loud quasars, while blue dashed contours represent the SDSS DR14 control sample. The top and right panels show the corresponding one-dimensional distributions of $\log M_{\rm BH}$ and $\log \lambda_{\rm Edd}$, respectively, with dashed lines indicating the median values.}
    \label{fig:Mbh_edd_desi}
\end{figure*}

Based on this DESI-parent-like SDSS control sample, we compared the emission line properties and black hole accretion properties of N-loud quasars with those of ordinary quasars. Figure~\ref{fig:contour_Civ_Mgii_desi} shows that, relative to the SDSS control sample, N-loud quasars tend to have narrower \Civ\; and \Mgii\; emission line widths, and their equivalent-width distributions are also systematically shifted. Figure~\ref{fig:Mbh_edd_desi} further shows that N-loud quasars are preferentially associated with lower black hole masses and higher Eddington ratios. 
Overall, the comparison shows that, relative to DESI-parent-like SDSS control sample, N-loud quasars show narrower broad emission lines, lower black hole masses, and higher Eddington ratios.


\bibliography{sample701}{}
\bibliographystyle{aasjournalv7}



\end{document}